\documentclass[fleqn,usenatbib]{mnras}

\usepackage{newtxtext,newtxmath}

\usepackage[T1]{fontenc}

\DeclareRobustCommand{\VAN}[3]{#2}
\let\VANthebibliography\thebibliography
\def\thebibliography{\DeclareRobustCommand{\VAN}[3]{##3}\VANthebibliography}

\usepackage{graphicx}	
\usepackage{amsmath}	

\title[The Cosmic Gamma-ray Horizon]{Mapping the Cosmic Gamma-ray Horizon:\\ The 1CGH Catalogue of Fermi-LAT detections above 10 GeV}

\author[B. Arsioli et al.]{ 
Bruno Arsioli,$^{1,2}$\thanks{E-mail: bsarsioli@ciencias.ulisboa.pt \& bruno.arsioli@gmail.com } 
Yu-Ling Chang,$^{3}$
Luca Ighina,$^{4,5}$
\\
$^{1}$Institute of Astrophysics and Space Science, OAL, University of Lisbon, Tapada da Ajuda, 1349-018 Lisbon, Portugal\\
$^{2}$Department of Physics, Faculty of Sciences, University of Lisbon, Edifício C8, Campo Grande, 1749-016 Lisbon, Portugal\\
$^{3}$Graduate Institute of Astrophysics, National Taiwan University, No. 1, Sec. 4, Roosevelt Road, Taipei 10617, Taiwan\\
$^{4}$Center for Astrophysics | Harvard \& Smithsonian, 60 Garden St., Cambridge, MA 02138, USA\\
$^{5}$INAF, Osservatorio Astronomico di Brera, via Brera 28, 20121, Milano, Italy\\
}

\pubyear{\the\year{2024}}

\begin{document}
\label{firstpage}
\pagerange{\pageref{firstpage}--\pageref{lastpage}}
\maketitle

\begin{abstract}
We present the First Cosmic Gamma-ray Horizon (1CGH) catalogue, featuring $\gamma$-ray detections above 10 GeV based on 16 years of observations with the Fermi Large Area Telescope (LAT) satellite. After carefully selecting a sample of blazars and blazar candidates from catalogues in the literature, we performed a binned likelihood analysis and identified 2791 $\gamma$-ray emitters above 10 GeV, at >3$\sigma$ level, including 62 that are new $\gamma$-ray detections. For each source, we estimated the mean energy of the highest-energy bin and analysed them in the context of the cosmic gamma-ray horizon. By adopting a reference model for the Extragalactic Background Light (EBL), we identified a subsample of 525 sources where moderate to severe $\gamma$-ray absorption could be detected across the redshift range of 0 to 3. This work provides the most up-to-date compilation of detections above 10 GeV, along with their redshift information. We condense extensive results from the literature, including reports on observational campaigns dedicated to blazars and $\gamma$-ray sources, thereby delivering an unprecedented review of the redshift information for sources detected above 10 GeV. Additionally, we highlight key 1CGH sources where redshift information remains incomplete, offering guidance for future optical observation campaigns. The 1CGH catalogue aims to track the most significant sources for studying the $\gamma$-ray transparency of the Universe. Furthermore, it provides a targeted subsample where the EBL optical depth, $\tau_{(E,z)}$, can be robustly measured using Fermi-LAT data.
\end{abstract}




\section{Introduction}

Blazars are a rare class of active galactic nuclei (AGN) characterised by relativistic jets pointing towards us, producing non-thermal emission that spans from radio to TeV $\gamma$-rays \citep{1995PASP..107..803U,2019ARA&A..57..467B}. Ranking among the most luminous and variable sources, blazars provide an unparalleled view into high-energy astrophysical processes, serving as cosmic beacons for studying the transparency of the Universe to $\gamma$-rays.

The Fermi Large Area Telescope \citep[Fermi-LAT,][]{FermiLAT_2009ApJ...697.1071A} has revolutionized our understanding of blazars, enabling us to collect data from tens of MeV to hundreds of GeV. Its vast energy coverage allows us to study the interaction of $\gamma$-rays with the Extragalactic Background Light (EBL), which leads to attenuation of the very high-energy (VHE) spectrum from distant sources \citep[e.g.][]{2013ApJS..209...34A,2018Sci...362.1031F}. This process is driven by photon-photon annihilation via electron-positron pair production, $\rm \gamma_{VHE} + \gamma_{EBL} \rightarrow e^{+} + e^{-}$ \citep[e.g.][]{osti_4836265,1967PhRv..155.1408G}, turning $\gamma$-rays blazars into unique probes of the EBL density and evolution over cosmic time \citep[e.g.][]{2013ApJ...771L..34D}. 

Direct EBL measurements are hindered by the presence of foregrounds, including zodiacal light \citep[e.g.][]{2022ApJ...941...33F}. However, distant extragalactic $\gamma$-ray sources, like blazars, provide indirect probes to map the EBL density and distribution over cosmic time \citep[e.g.][]{1992ApJ...390L..49S}. The EBL density in the far ultraviolet (FUV) to near-infrared (NIR) is largely driven by cumulative stellar activity along cosmic time \citep[e.g.][]{2001ARA&A..39..249H,2016ApJ...827....6S}. Thus, measuring the EBL at a given redshift provides an alternative diagnostic to recover the star formation rate (SFR) from earlier epochs, reaching as far back as the Epoch of Reionization \citep[e.g.][]{2010Natur.468...49R} at z$\sim$6. 

Given the sensitivity of Fermi-LAT, observations are particularly well suited to measuring the EBL in the UV to optical bands, whereas the near-infrared (NIR) can still be constrained at lower redshift. For example, \cite{2018Sci...362.1031F} demonstrated how distant $\gamma$-ray sources can trace the EBL density to z$\sim$3.0, providing unique insight into SFR close to the Epoch of Reionization (EoR).

The attenuation of $\gamma$-rays can be quantified by the EBL optical depth $\tau(E, z)$, which describes the cumulative probability of pair production and provides the relationship between observed and intrinsic flux: 

\begin{equation}
    F_{observed} = F_{intrinsic} \times \exp^{-\tau_{(E,z)}}
    \label{eq:abs}
\end{equation}
where a value of $\rm \tau_{(E,z)} = 1$ implies that approximately 63\% of the $\gamma$-ray flux at energy $E$ -from a source at redshift $z$- is absorbed. Measuring $\rm \tau_{(E, z)}$ is essential to understand the transparency of the Universe to gamma-rays, and provides a framework for testing and validation of EBL models. 

The attenuation of $\gamma$-rays by the EBL defines a statistical boundary for the farthest distance at which VHE photons of a given energy are likely to be detected, known as the \textit{Cosmic Gamma-ray Horizon} (CGH). The CGH concept was initially explored by \cite{osti_4836265,1967PhRv..155.1408G}, while \cite{1970Natur.226..135F} -with the Fazio Stecker relation- advanced the discussion of $\gamma$-ray attenuation in context of a cosmological origin for the isotropic $\gamma$-ray background.

Interestingly, the VHE spectra of distant blazars can exhibit an unexpected hardening at the highest energies \citep[e.g.][]{2010APh....33...81E,2015MNRAS.446.2267F}. This phenomenon has led to alternative hypotheses regarding the propagation of VHE photons, including photon oscillations into axion-like particles (ALPs) \citep[e.g.][]{2019MNRAS.487..123G,2020JCAP...09..027B,2024PDU....4401425A} and potential Lorentz invariance violation (LIV) \cite[e.g.][]{2020MNRAS.491.5268G,2024CQGra..41a5022A}. These mechanisms offer opportunities to probe fundamental physics and represent open questions in high-energy astrophysics. Therefore, a well-characterised sample of VHE emitters across a broad redshift range is essential, not only to probe EBL and SFR evolution more effectively, but also to investigate alternative photon propagation scenarios. 

In this work, we use Fermi-LAT observations spanning the first 16 years of the mission (August 2008 to August 2024) to search for $\gamma$-ray signatures associated with a sample of blazars and blazar candidates. By focusing on sources detectable above 10 GeV, we aim to provide a catalogue where the EBL absorption is most pronounce and could be effectively measured in the highest energy bins. 

To create a robust $\gamma$-ray sample suitable for studying EBL attenuation, we calculate the mean energy of the four highest-energy photons detected for each source. This approach offers a more stable and representative measure of the highest energy bin ($\rm E^{bin}_{max}$) detectable by Fermi-LAT, particularly when ranking sources that experience moderate to severe absorption, compared to using only the highest-energy photon ($\rm E_{max}$).

The First Cosmic Gamma-ray Horizon (1CGH) catalogue represents a significant step forward in describing the $\gamma$-ray horizon, and will contribute to our measurements of the EBL content and star formation rate across cosmic history. 

This work builds upon and complements existing efforts, such as the Third Fermi-LAT Catalog of High-Energy Sources \citep[3FHL;][]{2017ApJS..232...18A}, by adopting a strategy with the following key aspects: an extended observational period spanning 16 years, the targeted use of blazar positions as seeds for $\gamma$-ray analysis, the inclusion of sources down to a lower detection threshold (i.e., >3$\sigma$), and an extensive redshift review. These choices aim to highlight blazars that can serve as valuable targets for investigating the transparency of the Universe to $\gamma$-rays.

In addition, we propose a selection criterion that focuses on sources above the $\rm \tau_{(E, z)} > 0.1$ threshold, and build a subsample specifically targeted to the regime where at least 10\% of the $\gamma$-ray flux is expected to be absorbed in the highest energy bin.

An essential component of this work is the extensive redshift review of the detected sources. Initially, nearly half of them lacked redshift estimates from the reference catalogues (5BZcat, 3HSP, 3FHL, 4LAC-DR3). To address this gap, we conducted a comprehensive literature search, covering nearly 70 publications. We gathered information on spectroscopic redshift, lower limit estimates, and uncertain redshift (including photometric estimates reported in 4LAC-DR3); and introduce a redshift quality flag to differentiate between those cases. This effort significantly improved the reliability of our redshift data, providing a stronger foundation for subsequent studies.  

By providing a carefully curated sample of blazars with measurable EBL absorption, the 1CGH catalogue complements ongoing studies of TeV-detected sources \citep[e.g.][]{2023arXiv230400835G} and aims to deepen the scientific discussion on cosmic $\gamma$-ray propagation.

\section{Data Analysis and Methods}
\label{sec:data}

In this section, we discuss the selection of $\gamma$-ray candidates, the data analysis setup using the Fermi Science Tools, and the methods employed for cleaning and associating high-energy photons with sources detected above 10 GeV. We also outline the overall strategy used to ensure robust detections and photon associations, as reported in the 1CGH catalogue.

\subsection{Selection of Gamma-ray Candidate Sources}

To create a sample of sky positions and define our $\gamma$-ray targets, we combined several major blazar and AGN catalogues. We began by merging the `5th edition of the Roma-BZCAT Multi-Frequency Blazar Catalog' \citep[5BZcat;][]{2015Ap&SS.357...75M} with the `3rd edition of the High Synchrotron Peak sample' \citep[3HSP;][]{2019A&A...632A..77C}. These catalogues were chosen for their extensive blazar coverage and reliability.


Both 5BZcat and 3HSP are advanced versions of multifrequency catalogues containing thousands of blazars and blazar candidates, and represent significant repositories for the VHE community. These catalogues have been iteratively refined over time, building upon the initial BZcat \citep{2009A&A...495..691M}, and the First and Second Catalogues of High Synchrotron Peaked Blazars and Candidates, 1WHSP and 2WHSP \citep{2015A&A...579A..34A,2017A&A...598A..17C}.
   
We also included the latest version (V3.4) of TeVcat \citep{2008ICRC....3.1341W}, which contains a follow-up list of VHE-detected sources;  \url{http://tevcat.uchicago.edu}. Additionally, to improve the list of seed positions, we incorporate newly confirmed blazars and blazar candidates from recent literature, including \cite{2020ApJ...897..177P,2013ApJS..206...17M,2023MNRAS.519.2060I,2024A&A...692A.241I}.

We further include all sources classified as AGNs, blazars, and blazar candidates in the `Fourth Catalog of Active Galactic Nuclei Detected by the Fermi Large Area Telescope: Data Release 3' \citep[4LAC-DR3;][]{2022ApJS..263...24A}, since many of those candidates are not covered by the 5BZcat and 3HSP catalogues.

To remove duplicate entries, we use TopCat \citep{2005ASPC..347...29T} for internal matching; For the 5BZcat and 3HSP samples, the R.A. and Dec. positions, corresponding to the optical and infrared counterparts respectively, were used. For TeVcat and 4LAC-DR3 sources, we used the associated counterpart position whenever available (e.g., radio, infrared, or optical), as these typically offer better astrometric precision compared to the $\gamma$-ray position. With this sample of unique sources, we searched for $\gamma$-ray associations in the `Fermi LAT 14-Year Point Source Catalogue' \citep[4FGL-DR4;][]{2023arXiv230712546B} and the 3FHL catalogue \citep{2017ApJS..232...18A}, considering the 95\% containment region reported for each source.

At this point, we have successfully listed unique seed positions recovered from the 5BZcat, 3HSP, TeVcat and 4LAC-DR3 catalogues; But still, many 4FGL-DR4 high-latitude sources ($|b|>10^\circ$) were not included in our initial sample of 10 GeV candidates. This is partly due to new $\gamma$-ray detections from DR4 that had not yet been incorporated into the latest 4LAC release. However, the main reason is that many of these missing sources are actually unassociated in the 4FGL-DR4 --lacking classification. 

Therefore, in this work, we consider all unassociated high-latitude 4FGL-DR4 sources as valid 10 GeV candidates. Tracking detections above 10 GeV is essential, as it may motivate observational campaigns for their identification. 

In the final sample of seed positions, we removed all sources flagged as extended and as $\gamma$-ray bursts (GRBs). Finally, we used the \textit{CLASS1} flag from the 4FGL-DR4 catalogue to exclude cases associated with Galactic sources \citep[e.g., pwn, lmb, hmb, msp, glc, bin, nov, spp, snr,][]{2023arXiv230712546B}. The resulting sample contains nearly 8200 seed positions, with 5023 flagged as blazars or blazar candidates. The remainder are unknown types of sources from 4FGL-DR4 and TeVcat. Given how the positional seeds were selected, the 1CGH catalogue includes unassociated $\gamma$-ray sources that require further observational efforts to determine their nature, i.e, to confirm whether they are extragalactic in origin.

\begin{table*}
\caption{Power-law model parameters for the new $\gamma$-ray sources, with TS above 16, discovered in the 1CGH catalogue. A complete table with all 2791 1CGH sources is available in the online version of this paper (Note: A preliminary version is available in the Authors' GitHub repository). The first three columns show the original source names, right ascension (R.A.), and declination (Dec.) in degrees (J2000), respectively. The fourth column lists redshifts from literature, marked with a `(?)' flag for uncertain or photometric values, and with a `>' symbol for lower limits. The power-law parameters regarding the 1CGH analysis are reported in the following columns, corresponding to the fit (Equation \ref{eq:powerlaw}): the normalization `$\rm N_0$' in units of ph/cm$^2$/s/MeV, the photon spectral index `$\rm \Gamma$', and the flux integrated over 10-800 GeV, given in units of ph/cm$^2$/s. The pivot energy, `$\rm E_0$', is fixed at 15 GeV for all sources. The `TS' column indicates the Test Statistic value. The final column, $\rm E_{max}^{[GeV]}$, provides the highest energy photon associated with each 1CGH source, based on Pass 8 UltraClean event-class, with PSF=0 events removed, i.e. \texttt{evclass=512} and \texttt{evtype=56} (see Section \ref{sec:clean_photons}).}
\label{table:1CGH-16yrs}
{\def\arraystretch{1.3}
 \begin{tabular}{llrccccccc}
\hline
Name  & R.A. (deg)  &  Dec. (deg)  &  z  &  b(deg) & N$_0$ ($\times$10$^{-16}$) & $\rm \Gamma$ & Flux$^{10-800\, \rm GeV}_{( \times 10^{-12})}$  & TS  & E$^{[\rm GeV]}_{\rm max}$ \\
 \hline
  3HSPJ032356.5$-$010833 &  50.9855   & $-$1.14272    &  0.391   &  $-$45.15  &   37.30$\pm$6.90  &  2.43$\pm$0.25   &   69.7$\pm$11.9  &  167  &  50.5  \\ 
  3HSPJ120711.5$-$174605 &  181.79787 & $-$17.76830   &  >0.7    &   43.83    &   15.59$\pm$4.69  &  2.90$\pm$0.52   &   26.54$\pm$7.82 &  44.7 &  37.2  \\  
  3HSPJ220451.0$-$181536 &  331.21379 & $-$18.26011   &  0.26(?) &  $-$50.75  &    7.37$\pm$3.21  &  1.91$\pm$0.36   &   17.19$\pm$6.28 &  30.6 &  51.4  \\  
  3HSPJ030103.7+344101   &   45.26558 &    34.68366   &  0.246   &  $-$21.00  &    8.95$\pm$3.37  &  2.29$\pm$0.51   &   17.46$\pm$5.96 &  30.5 &  32.1  \\  
  3HSPJ132635.9+254958   &  201.64970 &    25.83300   &  0.698   &   82.02    &    4.46$\pm$2.33  &  1.62$\pm$0.35   &   12.88$\pm$4.97 &  28.9 &  116.8 \\  
  3HSPJ090802.2$-$095937 &  137.00921 &  $-$9.99369   &  0.054   &   24.38    &    5.41$\pm$1.04  &  1.80$\pm$0.14   &   13.50$\pm$2.70 &  26.7 &  70.6  \\   
  3HSPJ231023.4+311949   &  347.59737 &    31.33033   &  0.48(?) &  $-$26.77  &    5.30$\pm$0.31  &  1.85$\pm$0.11   &   12.88$\pm$1.15 &  26.1 &  51.8  \\  
  3HSPJ155424.1+201125   &  238.60054 &    20.19038   &  0.273   &   47.76    &   10.22$\pm$4.11  &  2.09$\pm$0.47   &   21.69$\pm$8.78 &  25.9 &  138.6 \\  
  3HSPJ101616.8+410812   &  154.07008 &    41.13680   &  0.27    &   55.32    &    4.04$\pm$2.22  &  1.77$\pm$0.39   &   10.39$\pm$4.41 &  24.1 &  164.2 \\  
  3HSPJ231041.8$-$434734 &  347.67400 & $-$43.79280   &  0.089   &  $-$63.75  &    3.86$\pm$0.30  &  1.71$\pm$0.10   &   10.37$\pm$0.94 &  23.6 &  48.7  \\  
  3HSPJ094606.1+215138   &  146.52554 &    21.86066   &  0.489   &   47.73    &    7.27$\pm$2.83  &  2.94$\pm$0.77   &   12.34$\pm$4.89 &  22.0 &  20.3  \\  
  3HSPJ091222.9$-$251825 &  138.09545 & $-$25.30702   &  0.33(?) &   15.62    &    7.76$\pm$3.04  &  2.45$\pm$0.50   &   14.41$\pm$6.00 &  21.8 &  52.1  \\  
  3HSPJ141003.9+051557   &  212.51633 &     5.26605   &  0.544   &   61.21    &    6.99$\pm$3.33  &  2.26$\pm$0.51   &   13.79$\pm$5.65 &  21.6 &  68.0  \\  
  3HSPJ061104.1+682956   &   92.76716 &    68.49905   &  0.5(?)  &   21.53    &    5.38$\pm$0.76  &  2.29$\pm$0.16   &   10.51$\pm$1.57 &  21.5 &  58.2  \\  
  3HSPJ024115.5$-$304140 &   40.31454 & $-$30.69447   &  0.3(?)  &  $-$65.76  &    6.73$\pm$1.06  &  2.25$\pm$0.23   &   13.32$\pm$2.33 &  21.3 &  59.9  \\  
  3HSPJ102523.0+040229   &  156.34595 &     4.04150   &  0.208   &   48.21    &    6.34$\pm$0.38  &  2.51$\pm$0.15   &   11.57$\pm$1.04 &  20.5 &  28.7  \\  
  3HSPJ121038.3$-$252713 &  182.65970 & $-$25.45383   &  0.47(?) &   36.50    &    8.60$\pm$1.95  &  3.22$\pm$0.47   &   14.31$\pm$3.25 &  19.7 &  26.3  \\  
  3HSPJ213004.8$-$563222 &  322.51987 & $-$56.53947   &  $-$     &  $-$43.91  &    5.26$\pm$2.47  &  1.94$\pm$0.39   &   12.06$\pm$4.80 &  18.4 &  57.2  \\  
  5BZQJ1353+0151         &  208.46492 &     1.86497   &  1.608   &   60.64    &    2.27$\pm$0.13  &  1.41$\pm$0.08   &   8.166$\pm$0.73 &  18.3 &  55.0  \\  
  3HSPJ094502.0$-$044833 &  146.25845 &  $-$4.80938   &  0.43(?) &   34.82    &    6.95$\pm$2.61  &  2.78$\pm$0.65   &   12.04$\pm$4.49 &  17.8 &  50.1  \\  
  3HSPJ120543.3+582933   &  181.43029 &    58.49266   &  0.4(?)  &   57.63    &    5.15$\pm$2.22  &  3.44$\pm$0.98   &   8.539$\pm$3.68 &  17.8 &  24.5  \\  
  3HSPJ213448.2$-$164205 &  323.70091 & $-$16.70155   &  0.8(?)  &  $-$43.51  &    7.97$\pm$3.48  &  3.63$\pm$1.19   &   13.23$\pm$5.74 &  17.3 &  29.5  \\  
  3HSPJ054903.0$-$215001 &   87.26416 & $-$21.83369   &  0.35(?) &  $-$23.05  &    5.12$\pm$2.45  &  2.46$\pm$0.66   &   9.495$\pm$4.17 &  17.2 &  66.5  \\  
  5BZGJ1840$-$7709       &  280.16083 & $-$77.15797   &  0.018   &  $-$25.80  &    5.97$\pm$2.57  &  2.44$\pm$0.52   &   11.10$\pm$4.54 &  16.8 &  29.8  \\  
  3HSPJ064326.7+421418   &  100.86133 &    42.23852   &  0.089   &   16.47    &    7.31$\pm$3.28  &  2.81$\pm$0.72   &   12.62$\pm$5.34 &  16.5 &  68.0  \\  
  3HSPJ100444.8+375212   &  151.18654 &    37.87000   &  0.44    &   53.58    &    4.52$\pm$0.65  &  3.86$\pm$0.47   &   7.580$\pm$1.11 &  16.3 &  13.7  \\  
  3HSPJ151136.9$-$165326 &  227.90375 & $-$16.89077   &  0.36(?) &   34.38    &    8.68$\pm$3.59  &  3.02$\pm$0.82   &   14.63$\pm$5.93 &  16.3 &  30.8  \\  
  3HSPJ104745.8+543741   &  161.94087 &    54.62813   &  0.54(?) &   54.46    &    4.56$\pm$2.07  &  2.61$\pm$0.71   &   8.165$\pm$3.64 &  16.3 &  31.7  \\  
  3HSPJ081941.8+053023   &  124.92433 &     5.50638   &  0.37(?) &   22.11    &    7.02$\pm$3.26  &  2.80$\pm$0.75   &   12.14$\pm$5.52 &  16.2 &  17.0  \\  
  3HSPJ062753.4$-$151957 &   96.97237 & $-$15.33252   &  0.3102  &  $-$12.04  &    5.32$\pm$2.89  &  1.85$\pm$0.41   &   12.89$\pm$5.63 &  16.2 &  38.8  \\  
  3HSPJ133326.0+623541   &  203.35820 &    62.59494   &  0.48(?) &   53.86    &    1.69$\pm$0.10  &  1.43$\pm$0.08   &   5.954$\pm$0.35 &  16.0 &  93.2  \\   
 \end{tabular}
}
\end{table*}

\subsection{Broadband Likelihood Analysis with Fermi-LAT Science Tools}
\label{sec:fermi-analysis}

To search for $\gamma$-ray signatures associated with the pre-selected seed positions, we performed a binned likelihood analysis in the 10-800 GeV band, covering the first 16 years of Fermi-LAT observations (August 2008 to August 2024). For the broadband analysis, we used the latest version of the \textit{Fermi Science Tools} (V2.2.0)\footnote{The Fermitools-Conda repository: \url{https://github.com/fermi-lat/Fermitools-conda}. The Fermi-LAT data analysis recommendations: \url{https://fermi.gsfc.nasa.gov/ssc/data/analysis/}.}, with Pass 8 event selection \citep[P8R3,][]{2013arXiv1303.3514A,2018arXiv181011394B}, following the Fermi-LAT's team recommendation for the identification of point-like sources. 

We worked with FRONT+BACK source-class events (\texttt{evtype=3} and \texttt{evclass=128}) and the instrument response function (IRF) P8R3-SOURCE-V3. For each analysis, we considered a region of interest (ROI) of 15$^\circ$ radius centred on the $\gamma$-ray seed positions. For modelling the background of point and extended sources, we adopted the 4FGL-DR4 v34 catalogue (\textit{gll-psc-v34.fit}), along with isotropic and galactic-diffuse templates (\textit{iso-P8R3-SOURCE-V3-v1.txt} and \textit{gll-iem-uw1216-v13.fits}, respectively). The spectral parameters --normalization and photon index-- were set free for all sources within 5$^\circ$ of the seed position.

A zenith angle cut of 105$^\circ$ was applied to avoid contamination from Earth's limb $\gamma$-ray photons, which are produced by cosmic-ray interactions with the atmosphere. Using the \texttt{gtmktime} routine, we selected good time intervals when Fermi-LAT was operating in `science data-taking mode,' by setting the flags \texttt{DATA-QUAL$>$0} and \texttt{LAT-CONFIG==1}. Using the \texttt{gtbin} routine, we generated counts maps (CMAP) and counts cubes (CCUBE) with dimensions of $300 \times 300$ and $200 \times 200$ pixels at $0.1^\circ$/pixel, respectively. For the CCUBE, we used 20 equally-spaced logarithmic energy bins in the 10-800 GeV range. Each $\gamma$-ray candidate was modelled as a point-like source characterised by a power-law spectrum:
\begin{equation} \rm
\hspace{85pt}\frac{dN}{dE} \, \text{=} \, N_0 \; \left( \frac{E}{E_0} \right)^{-\Gamma} \;,
\label{eq:powerlaw}
\end{equation} 

In this equation, $\rm N_0$ is the normalization constant (prefactor), given in units of [photons/cm$\rm ^2$/s/MeV], representing the flux density at the pivot energy $\rm E_{0}$. The parameter $\rm \Gamma$ represents the photon spectral index, indicating the slope of the spectrum. To restrict the parameter space probed during the fitting process, we set a limit of 6.5 for the maximum $\rm \Gamma$ value. This limit is compatible with the largest spectral indices observed in the 3FHL catalogue, including all source types \citep[i.e. Galactic, extragalactic, and unassociated/unknown sources,][]{2017ApJS..232...18A}

In our analysis, we adopt a low-energy threshold of 10 GeV --similar to the 3FHL catalogue-- due to the improved angular resolution and lower background contamination compared to lower thresholds \citep{2017ApJS..232...18A}. For the upper energy limit, we set a conservative threshold of 800 GeV, whereas 3FHL extended the analysis to 2 TeV. This decision follows from the discussion in the 4FGL paper \citep[Section 3.2 of][]{2020ApJS..247...33A}, which highlights that a broadband analysis reaching 1 TeV can introduce uncertainties in the energy flux estimate, particularly for hard-spectrum sources. Therefore, our choice for the high-energy threshold is meant to mitigate uncertainties regarding the spectral fitting of all sources in the ROI region.

\subsection{Detection Threshold for Multifrequency-Selected Targets}
\label{sec:detect_threshold}

The use of multifrequency seed positions to detect $\gamma$-ray sources has been successfully applied in numerous works \citep[][]{2017A&A...598A.134A,2018MNRAS.480.2165A,2018A&A...616A..20A,2020MNRAS.493.2438A} and has proven effective in uncovering extreme blazars close to the Fermi-LAT detection threshold. Using this approach, the analysis for each sky position is independent, investigating a single $\gamma$-ray candidate at a time. To handle all 8200 candidates efficiently, we relied on parallel High Performance Computing (HPC) resources provided by INCD-Portugal.  

Considering our likelihood analysis involves only two degrees of freedom ($\rm \Gamma$ and $\rm N_{0}$) we adopted a pre-selection threshold\footnote{The Test Statistic (TS) parameter is defined as $ \rm -2 \ln \left( L_{\left( no-source \right)} \div L_{\left( source \right)} \right) $, where $\rm L_{ \left( no-source \right)}$ represents the likelihood of observing a given photon count due to background alone (the null hypothesis), and $\rm L_{\left( source \right)}$ represents the likelihood of observing the photon count assuming a source exists at a particular position \citep{1996ApJ...461..396M}.} of TS > 12 (equivalent to a 3.03$\sigma$ detection\footnote{See the repository \textit{TS-DegFreedom-Sigma-relation-FermiLAT} at \url{github.com/BrunoArsioli} with a python implementation for the TS to sigma relation, while accounting for the number of degrees of freedom in the analysis.}) to include faint sources for future follow-up observations. As suggested in \cite{1996ApJ...461..396M},  the use of multifrequency seeds improves the $\gamma$-ray detectability of point-like sources near the detection threshold of high-energy observatories, making it a robust strategy for data exploration; especially regarding blazars, which are known to be the dominant extragalactic population of GeV emitters.

In addition, this relatively low selection threshold aims not only to highlight faint sources but also to capture those that undergo short-lived flare episodes, which might otherwise be missed over a long integration period due to signal dilution \citep{2018A&A...616A..20A}. As the true signal (i.e. the photon counts from a transient flare) remains constant while the background noise accumulates, a lower TS threshold allows us to include those valuable cases. In short time windows, the significance of these flaring events would be higher, enabling the study of the effects of the EBL absorption. Although we did not perform light-curve analyses in this work, the 1CGH catalogue can serve as a basis for future time-resolved studies.

To assess the robustness of our detection threshold, we estimated the spurious detection rate in our analysis setup. In a chi-squared ($\chi^2$) distribution with two degrees of freedom (d.o.f), the cumulative probability $ P(TS>TS_{threshold})$ represents the likelihood of obtaining a test statistic above the threshold purely by chance. Using TS=12, the spurious detection probability is calculated as $\rm p\_value = 1  - \chi^2 \times cdf(TS=12, \  d.o.f=2) \approx 0.0025$, where $cdf()$ is the cumulative distribution function. Given that 8200 seed positions were analysed, the expected number of spurious detections in the pre-selection phase is approximately $\rm 8200\times0.0025 = 20.5$. This value represents an upper limit of contamination. 

In next section, we discuss an additional selection criteria --beyond the selection based on TS level alone-- to avoid spurious detections and improve reliability of the final catalogue.

\subsection{Additional Selection Criteria: Refining the 1CGH Sample}
\label{sec:clean_photons}

As a result of the likelihood analysis, we pre-selected 3004 sources with TS > 12, which are used as a base for building the final sample. Those sources will still go through an additional selection step that involves looking at the 
$\gamma$-ray events associated with each source. In this stage, it is essential to consider the different event types in the dataset and also to identify and remove potential sources of contamination from the event sample.

In the Fermi-LAT Pass 8 release \citep{2013arXiv1303.3514A}, each event is assigned a Point Spread Function (PSF) type, which indicates the quality of the reconstructed direction. Events are categorized into four quartiles: PSF0 (lowest quality, \texttt{evtype=4}), PSF1 (\texttt{evtype=8}), PSF2 (\texttt{evtype=16}), and PSF3 (highest quality, \texttt{evtype=32}). The PSF 68\% containment radius varies by energy and event type. For example, at 30 GeV, the containment radius for PSF1 events is approximately 0.12$^\circ$, as described in the Pass 8 documentation\footnote{Fermi-LAT Pass 8 PSF: \url{https://www.slac.stanford.edu/exp/glast/groups/canda/lat_Performance.htm}}.

Using the \texttt{gtselect} routine, we created a subsample of the all-sky Fermi-LAT data, selecting only events with energy greater than 10 GeV. We removed low-quality events by excluding PSF0 events, which represent the quartile with the poorest reconstruction accuracy. Next, we used the \texttt{gtmktime} routine to select events during good time intervals (see Section \ref{sec:fermi-analysis}). After this step, we were left with a photon sample containing PSF1-2-3 source-type events (i.e., \texttt{evclass=128} and \texttt{evtype=56}), with a maximum zenith angle of 105$^\circ$, all recorded under `science data-taking mode'.

Before recovering the $\gamma$-ray events associated with the pre-selected sources, we applied an additional layer of data cleaning to improve the quality of the photon sample. Specifically, we removed events associated with solar emissions and $\gamma$-ray bursts (GRBs). Indeed the solar disk has recently been confirmed as a source of very high-energy $\gamma$-rays \citep{2022PhRvD.105f3013L,2023PhRvL.131e1201A,2024ApJ...962...52A}; Therefore, we removed all events that might be linked to solar emissions by assuming an association radius of 0.8$^\circ$\footnote{Considering the solar disk has an angular radius of $\sim$0.26$^\circ$, and the characteristic position uncertainty for PSF1 events at 10 GeV is around 0.18$^\circ$, a 0.8$^\circ$ radius cut safely encompasses and removes solar disk contamination.}.

To identify and remove events related to GRBs, we used the Fermi GBM Burst Catalogue \footnote{The continuously updated GBM Burst Catalogue, also referred to as 'The FERMIGBRST database', is available at \url{https://heasarc.gsfc.nasa.gov/W3Browse/fermi/fermigbrst.html}.} \citep{2020ApJ...893...46V}. We removed events recorded within 0.8$^\circ$ of GRB locations, also considering a time window of 30 minutes before and 10 hours after each GRB.

The characteristic position uncertainty of PSF1 events at 30 GeV was used as a basis for defining a cross-matching radius of 0.12$^\circ$ between the pre-selected candidates and the clean photon sample. This conservative radius ensures robustness when associating higher-energy events, particularly given the larger PSF1 size at the lowest energy (approximately 0.18$^\circ$ at 10 GeV).

As an intermediary selection layer of the 1CGH catalogue, we retained only pre-selected sources (TS>12) that were associated with at least four high-energy events from the clean photon sample, matching within 0.12$^\circ$. This approach follows similar criteria used in building the 3FHL sample \citep{2017ApJS..232...18A}, which required a minimum of four associated photons for source acceptance (as predicted by the adjusted model, i.e., $N_{pred} \geq 4$). This procedure removed 213 sources ($\rm \sim7\%$) from the 3004 pre-selected ones. We are left with a final sample of 2791 sources, and --given the selection criteria described above-- the expected level of spurious detections should be lower than the 0.68\% figure (i.e., 20.5/3004) presented in Section \ref{sec:detect_threshold}.


\section{The First Cosmic Gamma-ray Horizon Catalogue (1CGH)}

We present the First Cosmic Gamma-ray Horizon catalogue (1CGH), which lists 2791 blazars and blazar candidates detected with a TS greater than 12 in the 10-800 GeV energy range, integrated over 16 years of Fermi-LAT observations. For detailed information on the catalogue's metadata, refer to Table \ref{tab:1CGHcolumns} in Appendix \ref{appendix:1CGHcolumns}. 

The 1CGH catalogue includes 62 $\gamma$-ray detections never reported in earlier Fermi-LAT catalogue releases (1FGL, 2FGL, 3FGL, 4FGL, 2FHL and 3FHL), representing new $\gamma$-ray sources. These newly identified sources are mostly high synchrotron peak blazars from the 3HSP catalogue \citep{2019A&A...632A..77C}, and could be especially significant as candidates for VHE observations with the upcoming Cherenkov Telescope Array Observatory \citep[CTAO,][]{2019scta.book.....C,2013APh....43....3A}. The 3HSP catalogue was designed to identify VHE targets for CTAO, and here we confirm its potential.

Table \ref{table:1CGH-16yrs} presents a selection of newly detected sources with TS above 16, including the seed-source name, right ascension (R.A.), declination (Dec.), redshift, and power-law model parameters associated with each new $\gamma$-ray detection\footnote{The full version of the First Cosmic Gamma-ray Horizon catalogue will be available in public data repositories such as Vizier \url{https://vizier.cds.unistra.fr/viz-bin/VizieR} and Author's GitHub.}. 

\begin{figure}
	\includegraphics[width=1.0\linewidth]{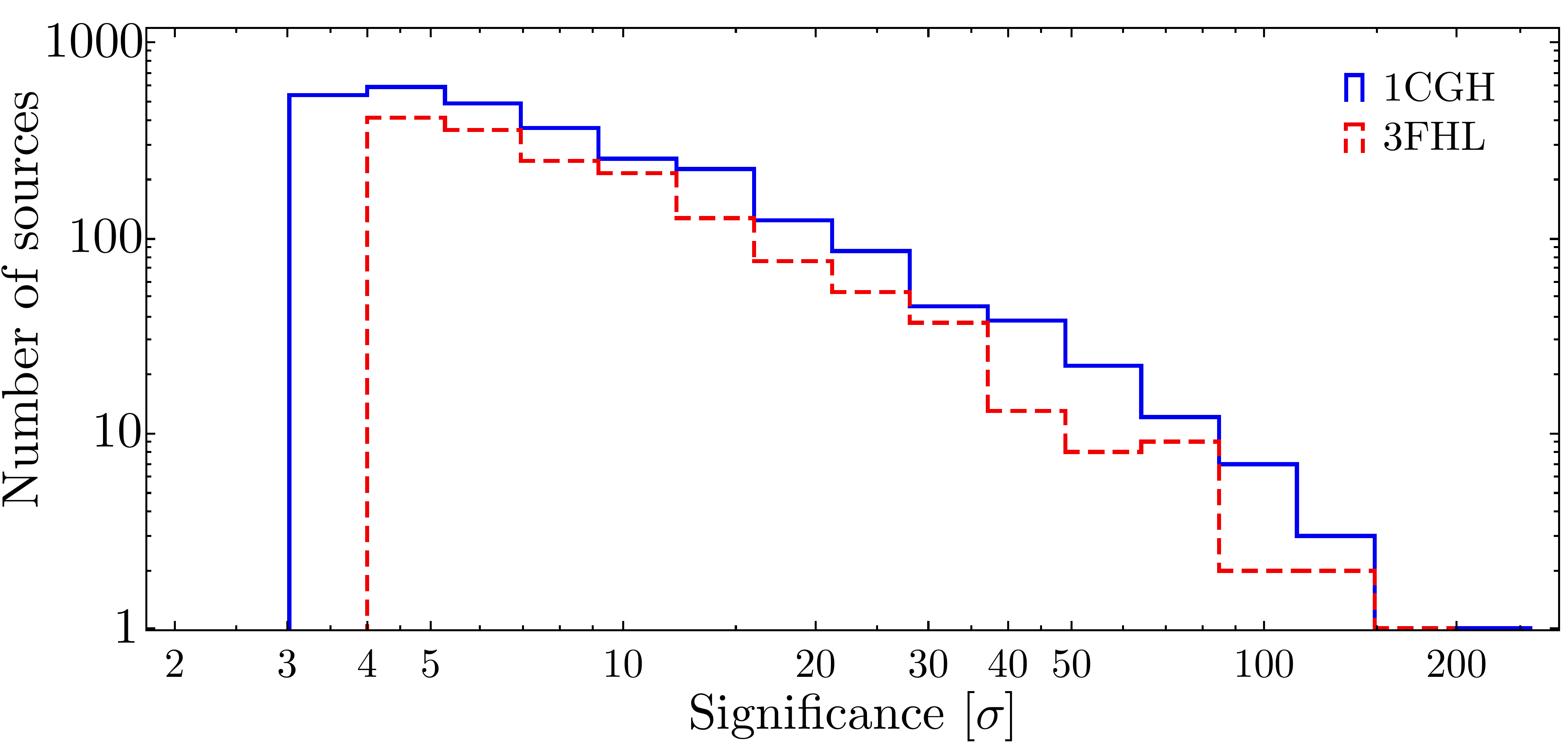}
    \caption{The distribution of detection significance ($\sigma$) for the 1CGH and 3FHL samples (respectively, blue line and red-dashed line).}
    \label{fig:sigma}
\end{figure}

In Fig. \ref{fig:sigma} we show the distribution of the detection significance for the entire 1CHG and 3FHL catalogues. This representation is meant as a qualitative overview on both samples, keeping in mind that the 1CGH is focus on seed-positions of blazars and blazar-candidates, while the 3FHL includes detections of all source types. The number of detections in 1CGH shows consistent growth along the significance-bins (i.e. following a trend similar to 3FHL), and includes an extra bin covering the 3<$\sigma$<4 range.

We should note that in the full 1CGH catalogue, 26 sources\footnote{Comprising 3 unassociated sources, 4 BL Lacertae, 8 Flat Spectrum Radio Quasars (FSRQs), and 11 blazars of unknown type.} reached the spectral index limit of $\Gamma = 6.5$ constrained by our likelihood analysis setup (see Section \ref{sec:fermi-analysis}). Moreover, 27 1CGH sources have counterparts in 3FHL where spectral curvature has been identified\footnote{To help identify such cases, we include the \textit{Curvature\_3FHL} column to our final catalogue, which tracks whether spectral curvature was identified in 3FHL, and might affect broadband flux estimates.}. For these 53 (26+27) cases, the derived fluxes should be interpreted as upper limits. While this affects only a small fraction of the sample, the vast majority of sources exhibit well-constrained spectral fits, reinforcing the robustness of the 1CGH catalogue for high-energy studies.

\begin{figure}
	\includegraphics[width=1.0\linewidth]{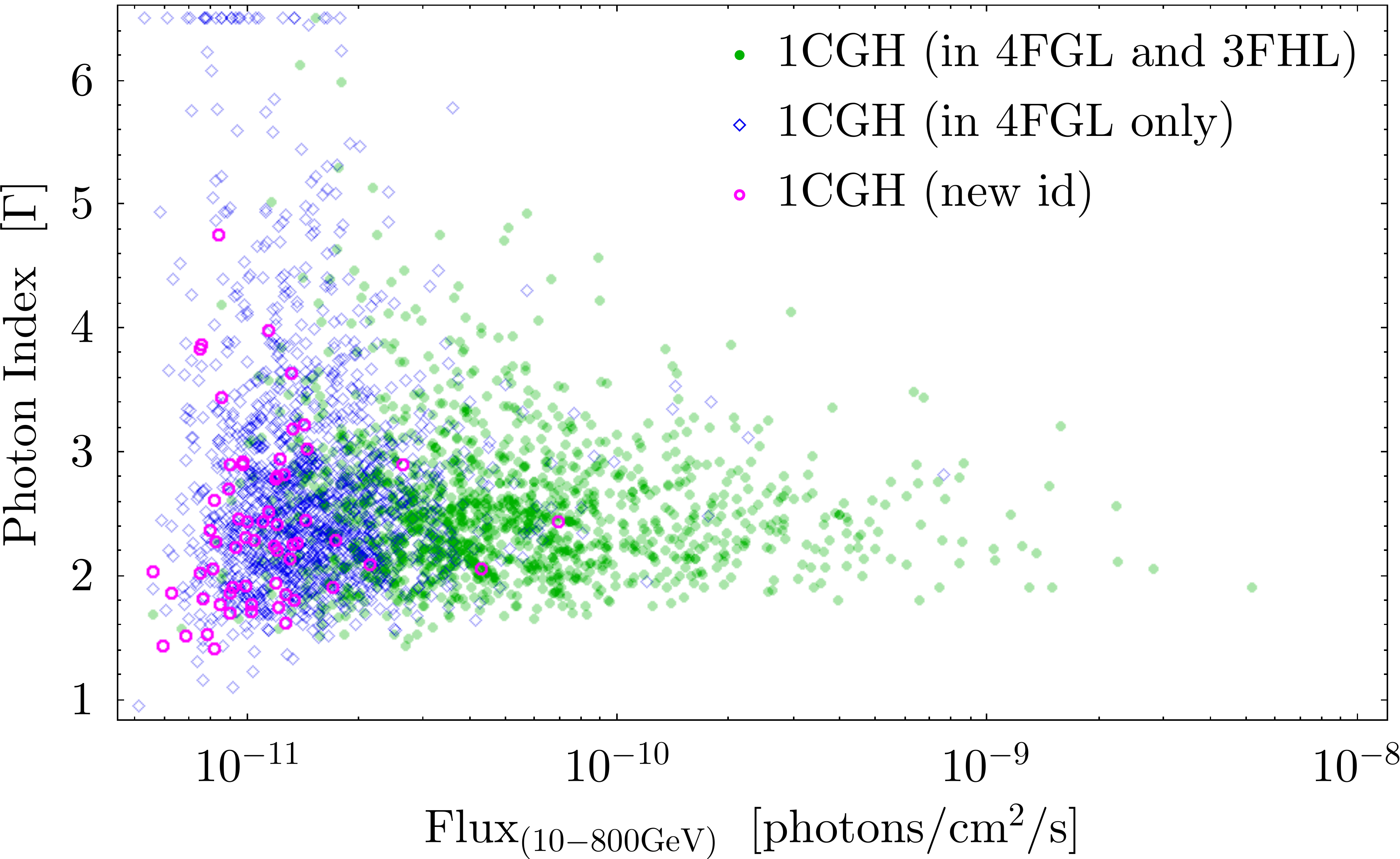}
    \caption{The photon index ($\Gamma$) versus the integral flux in the 10 GeV to 800 GeV energy band. The green points represent 1CGH sources with counterparts in both 3FHL and 4FGL-DR4 catalogues; the blue open diamonds represents 1CGH sources with counterparts in 4FGL-DR4 only; and the magenta circles represents new $\gamma$-ray identifications.}
    \label{fig:index_vs_flux}
\end{figure}

In Fig. \ref{fig:index_vs_flux} we present the photon index ($\Gamma$) versus the integrated flux derived from a power-law fit over the 10–800 GeV energy range. The plot shows that the 1CGH sample extends the 3FHL coverage toward fainter fluxes, effectively lowering the sensitivity limit for detections above 10 GeV. As in the 3FHL catalogue, the detection threshold in the 1CGH sample shows little dependence on photon index, with the new identifications clustering at the lowest fluxes. Moreover, the overlap between the 1CGH and 4FGL sources highlights a robust improvement in completeness above 10 GeV, which is complemented by the newly identified sources.

\subsection{Extensive Redshift Review}
\label{sec:redshift}

In addition to listing detections above 10 GeV, we conducted an extensive literature review to improve the redshift characterization of our sample, which is crucial for investigating the cosmic $\gamma$-ray horizon. Based on the latest studies, we revised and updated redshift information, tracked the origin and quality of the data, and incorporated new redshifts from multiple optical campaigns dedicated to characterizing blazars, 3FHL and 4FGL sources. To track redshift quality, we added a column named `z-flag' to the catalogue, with values: (1) for robust spectroscopic redshift; (2) for photometric estimates or uncertain values reported in literature; and (3) for lower-limit redshift values.

In the 1CGH catalogue, we added a column named `z-origin' to track the origin of the redshift values reported in the `z' column. Redshift information was gathered from several major blazar and $\gamma$-ray catalogues, including 5BZcat \citep{2015Ap&SS.357...75M}, 3HSPcat \citep{2019A&A...632A..77C}, 4LACdr3 \citep{2022ApJS..263...24A}, and TeVcat \citep{2008ICRC....3.1341W}. 

Additionally, redshift values were extracted from dedicated observational campaigns, such as Boyle90 \citep{1990MNRAS.243....1B}, Appenzeller98 \citep{1998ApJS..117..319A}; Sbarufatti05 \citep{2005AJ....129..559S}; Sbarufatti06 \citep{2006A&A...457...35S}; Sbarufatti09 \citep{2009AJ....137..337S}; Maisner10 \citep{2010ApJ...712...14M}; Pulido10 \citep{2010A&A...519A...5A}; Shaw12 \citep{2012ApJ...748...49S}; Furniss13 \citep{2013ApJ...766...35F}; Landoni13 \citep{2013AJ....145..114L}; Maselli13 \citep{2013ApJS..206...17M}; Rovero13 \citep{2013ICRC...33.2676R}; Sadrinelli13 \citep{2013AJ....146..163S}; Shaw13 \citep{2013ApJ...764..135S}; Paggi14 \citep{2014AJ....147..112P}; Ricci15 \citep{2015AJ....149..160R}; AlvarezCrespo16a-b-c \citep{2016AJ....151...32A,2016AJ....151...95A,2016Ap&SS.361..316A}; Kaur16 \citep{2017ApJ...834...41K}; Rovero16 \citep{2015arXiv150908377R,2016A&A...589A..92R}; Juanita17 \citep{2018MNRAS.474.3162T}; Paiano17 \citep{2017ApJ...851..135P,2017ApJ...837..144P,2017ApJ...844..120P}; Gabanyi18 \citep{2018A&A...612A.109G}; Landoni18 \citep{2015AJ....149..163L,2018ApJ...861..130L}; Massaro18 \citep{2014AJ....148...66M,2015A&A...575A.124M,2016Ap&SS.361..337M}; Mishra18 \citep{2018MNRAS.473.5154M}; Balmaverde19 \citep{2020MNRAS.492.3728B}; Caccianiga19 \citep{2019MNRAS.484..204C}; Johnson19 \citep{2019ApJ...884L..31J}; Desai19 \citep{2019ApJS..241....5D}; Menezes19 \citep{2019A&A...630A..55D,2020Ap&SS.365...12D}; Marchesini19 \citep{2019Ap&SS.364....5M}; Paiano19 \citep{2019ApJ...871..162P}; Belladitta20 \citep{2020A&A...635L...7B}; Goldoni20 \citep{2015ICRC...34..835G,2021A&A...650A.106G}; Landoni20\footnote{ZBLLac Database: ``A spectroscopic Library of BL Lac objects" at \url{https://web.oapd.inaf.it/zbllac/}} \citep{2020ApJS..250...37L}; Pena-Herazo20 \citep{2020A&A...643A.103P,2019Ap&SS.364...85P,2017Ap&SS.362..228P}; Paiano20 \citep{2019ApJ...871..162P,2020MNRAS.497...94P}; Raiteri20 \citep{2020MNRAS.493.2793R}; B.Gonzalez21 \citep{2021MNRAS.504.5258B}; Paliya21 \citep{2021ApJS..253...46P}; Pena-Herazo21 \citep{2021AJ....162..177P,2021AJ....161..196P}; Rajagopal21 \citep{2021ApJS..254...26R}; Foschini22 \citep{2022Univ....8..587F}; Kasai22 \citep{2023MNRAS.518.2675K,2023IAUS..375...96K}; OlmoGarcia22 \citep{2022MNRAS.516.5702O}; Rajagopal22 \citep{2023AJ....165...42R}; Paiano23 \citep{2023MNRAS.521.2270P}; dAmmando24 \citep{2012MNRAS.427..893D,2024A&A...683A.222D}; Sarira24 \citep{2024MNRAS.533.2156S}, DESI-EDR \citep{2024AJ....168...58D}, and AlvarezCrespo25 \citep{2025A&A...694A..46A}. The numerous dedicated optical observation campaigns highlight the ongoing high demand for the characterization of blazars and blazar candidates, particularly in connection with their $\gamma$-ray counterparts.

Starting from the redshift information available in the reference catalogues (i.e., 5BZcat, 3HSP, 4LAC-DR3, and TeVcat), our literature review allowed us to assign or update redshift values for 967 1CGH sources --including 377, 437, and 153 sources with z-flags 1, 2, and 3, respectively. Of the total 2791 1CGH sources, 1062 (38.2\%) have spectroscopically measured redshift values, 855 (30.6\%) have uncertain/photometric values, 210 (7.5\%) have lower-limit values, and 664 (23.7\%) have no available redshift information.

\begin{figure}
	\includegraphics[width=1.0\linewidth]{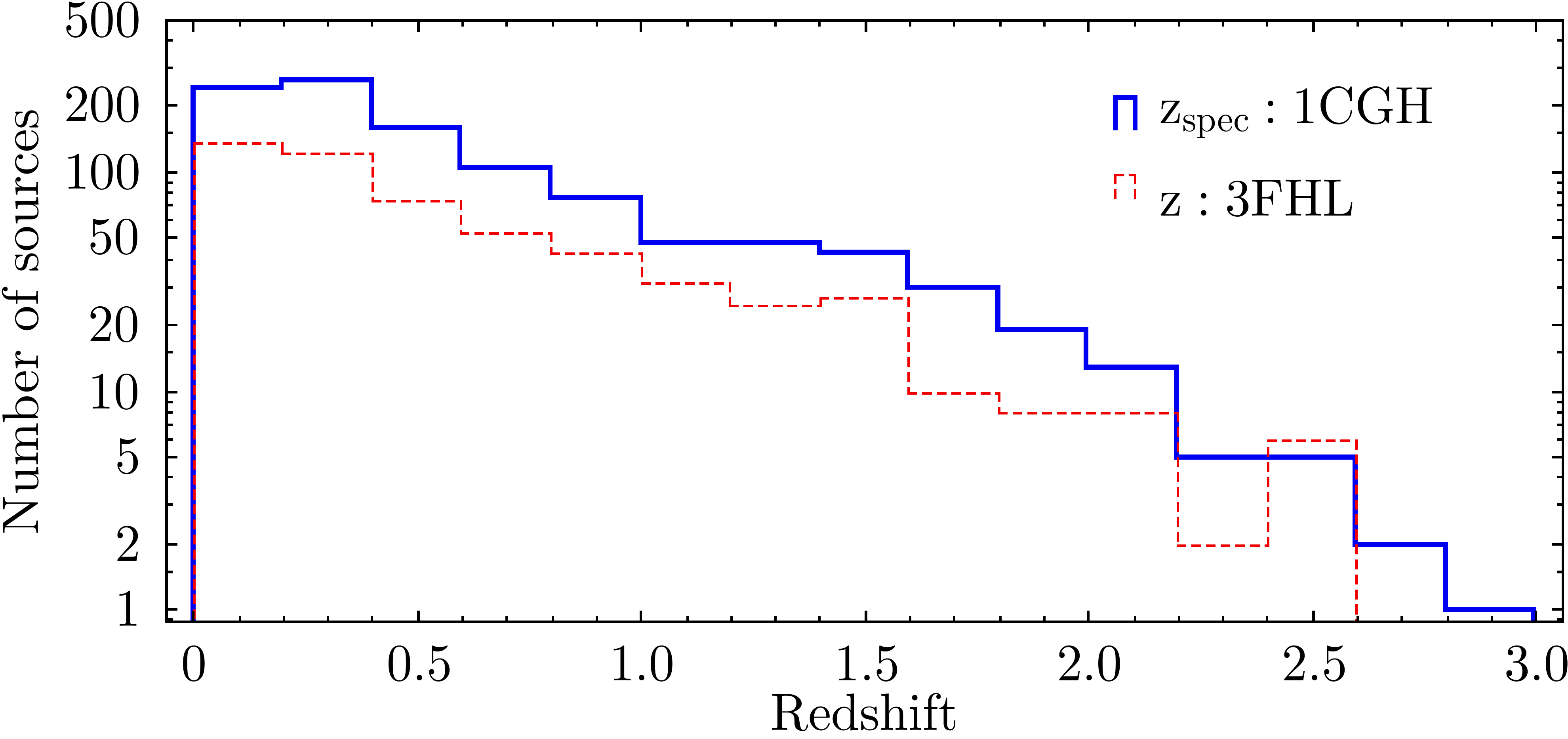}
    \caption{The redshift distribution for the 1CGH and 3FHL samples. For the 1CGH, the histogram only include sources with spectroscopic redshift determination (\textit{z-flag}=1); for the 3FHL catalogue, we include the entire sample.}
    \label{fig:zhisto}
\end{figure}

In Figure \ref{fig:zhisto}, we show the redshift distribution for the 1CGH subsample with z-flag=1 (i.e. robust spectroscopic redshift) and for the entire 3FHL sample. The comparison reveals that the number of sources detected above 10 GeV has improved along redshift relative to 3FHL. In addition, our redshift characterization aims to complement and refine previous efforts (e.g., 3FHL and 4LAC-DR3) by incorporating redshift quality flags and systematically tracking the reference from which each redshift value was obtained\footnote{Please refer to Sec. 5.2 of \cite{2020ApJ...892..105A} regarding redshift contamination in 4LAC, given the lack of flags to differentiate between spectroscopic and photometric redshift values.}.  

Extensive redshift characterization is key for samples used to study the CGH, as it helps to mitigate possible biases caused by missing or uncertain redshift values. The best way to reduce uncertainties when measuring the EBL density evolution is to increase the number of sources with robust redshift determinations. In particular, identifying absorbed sources above redshift 2 would significantly improve EBL measurement at a regime that impacts SFR density estimates near and within the EoR, at z$\sim$6-7 \cite{2018Sci...362.1031F}. Among the most distant 1CGH sources (at $\rm 2.4\leq z \leq2.9$) are: 5BZQJ0601-7036, 5BZQJ0228-5546, 5BZQJ1344-1723, 5BZQJ1748+3404, 5BZQJ1441-1523, 5BZQJ0912+4126, 5BZQJ1345+4452 and 5BZBJ0022+0608, which could contribute to studies of the EoR.

With improved precision in EBL density measurements, we can potentially disentangle the contributions of star formation and AGN to the EBL build-up at high redshift, turning blazars into an effective tool to understand the role played by AGNs during the EoR \citep[e.g.,][]{2023ApJ...959L..18D}. 

\subsection{The Cosmic Gamma-ray Horizon Plot}

\begin{figure*}
	\includegraphics[width=1.0\linewidth]{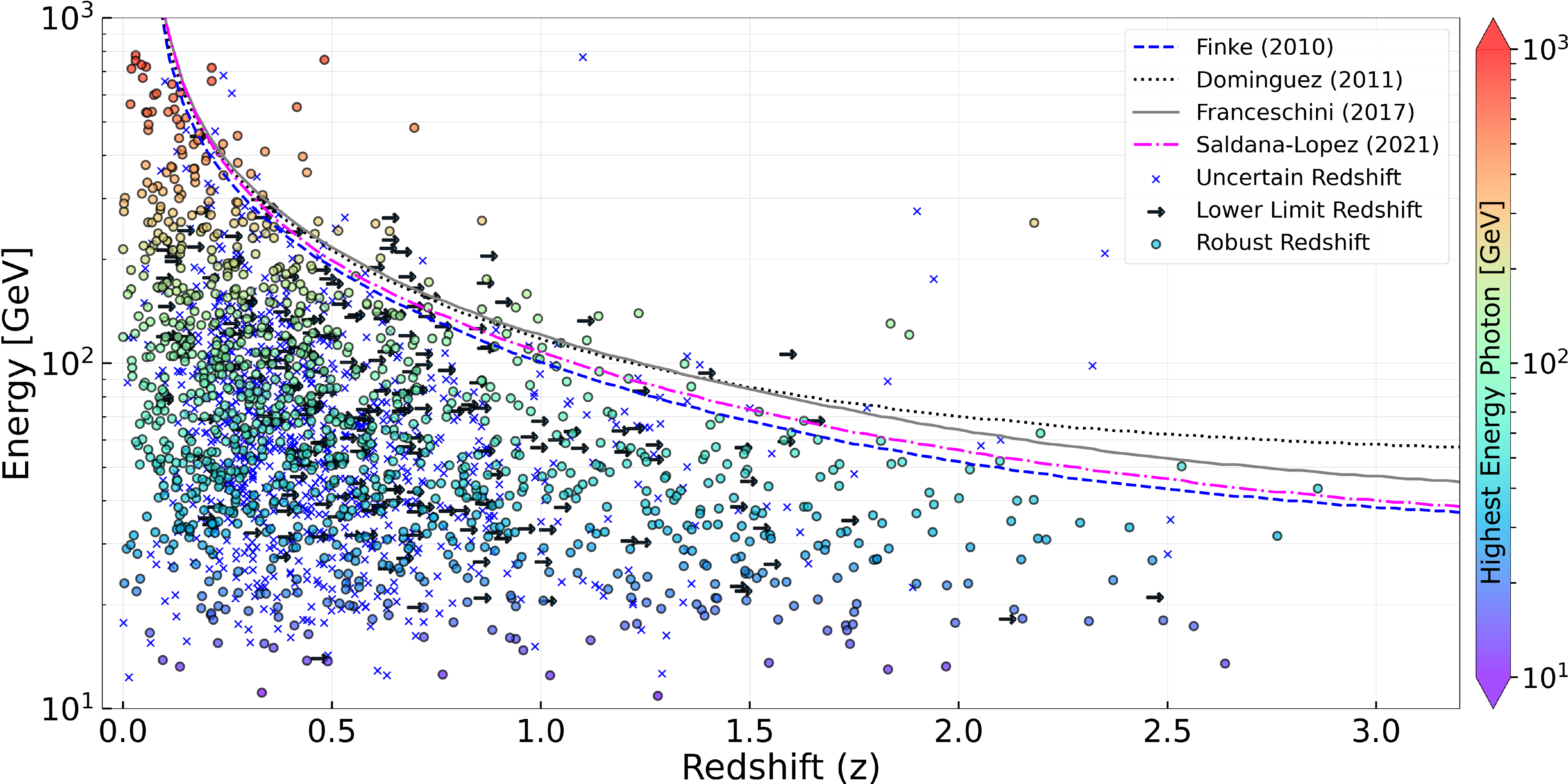}
    \caption{The cosmic gamma-ray horizon, showing the highest-energy photon versus redshift, based on Fermi-LAT ultra-clean events with PSF0 events removed (\texttt{evclass=512} and \texttt{evtype=56}). Sources detected above 10 GeV are represented, with robust (spectroscopic) redshifts marked as filled circles; lower-limit redshifts are represented by right arrows, with the reported redshift value corresponding to the center of the arrow}; and uncertain redshifts (photometric or reported as doubtful) indicated by blue crosses. The EBL optical depth $\rm \tau_{(E,z)} = 1.0$ is depicted for different models: \citet{2010ApJ...712..238F} (blue dashed line), \citet{2011MNRAS.410.2556D} (black dotted line), \citet{2017A&A...603A..34F} (grey line), and \citet{2021MNRAS.507.5144S} (magenta dot-dashed line). For better visualization, the photon energy of sources with robust redshift is colour-coded. Note that approximately one-third of the 1CGH sources lack assigned redshift and are therefore not included in this plot.
    \label{fig:cgh}
\end{figure*}

Plotting the highest-energy photons from the 1CGH catalogue against redshift reveals that the Universe is indeed opaque to $\gamma$-rays, as predicted by \cite{osti_4836265,1967PhRv..155.1408G}. As previously discussed, the energy vs. redshift relationship, where the opacity $\rm \tau(E,z) = 1.0$, defines what is known as the `Cosmic Gamma-Ray Horizon'. Beyond this horizon, high-energy photons are severely attenuated by the EBL, rendering the Universe effectively opaque to $\gamma$-rays \citep[e.g.][]{2013ApJ...770...77D}. 

Figure \ref{fig:cgh} shows the highest-energy photon versus redshift for the 1CGH sources, along with the predicted gamma-ray horizon according to the main EBL models \citep[e.g.][]{2010ApJ...712..238F,2011MNRAS.410.2556D,2017A&A...603A..34F,2021MNRAS.507.5144S}. Different markers are used to represent robust (spectroscopic), photometric/uncertain, and lower limit redshift values to emphasize the importance of accurate redshift determination. 

Figure \ref{fig:cgh} provides an unprecedented view of the redshift characterization of sources used to study the CGH. It highlights the significance of reaching an extensive redshift description of the entire 1CGH sample and aims to further motivate the community's ongoing efforts in the optical characterization of these rare sources (see Section \ref{sec:redshift}). 

Here we draw attention to the number of 1CGH sources that lack robust redshift characterization; As mentioned in Section \ref{sec:redshift}, a total of 1729 1CGH sources ($\sim$72\%) are assigned redshift values that are either uncertain, lower limits, or are absent. All of those cases would benefit from dedicated observational campaigns to improve the overall redshift description, therefore representing a significant challenge for CGH studies and the very high-energy community as a whole.

\subsection{Identification of Sources with Detectable EBL Absorption}

\begin{figure*}
	\includegraphics[width=1.0\linewidth]{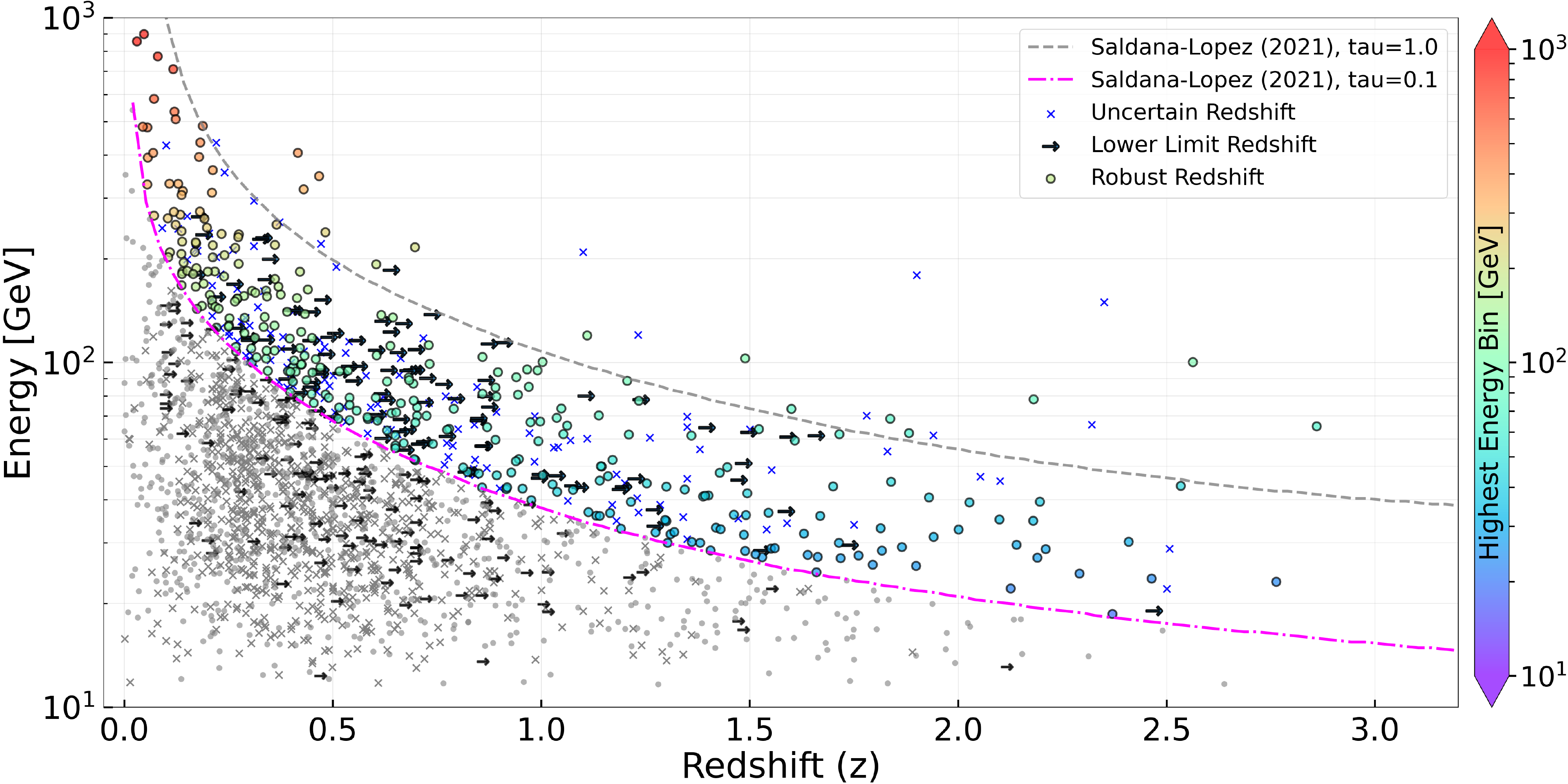}
    \caption{The cosmic gamma-ray horizon, showing the `largest energy bin detectable with Fermi-LAT' versus redshift. The `largest energy bin' is calculated as the mean energy of the four highest-energy source-type events associated with each source, excluding PSF0 events (i.e., \texttt{evclass=128} and \texttt{evtype=56}). The opacity regimes for EBL optical depths $\rm \tau_{(E,z)} = 1.0$ and $\rm \tau_{(E,z)} = 0.1$ are represented by grey dashed and magenta dot-dashed lines, respectively, corresponding to the Saldana-Lopez model \citep{2021MNRAS.507.5144S}. This plot highlights sources in the $\rm \tau_{(E,z)} > 0.1$ regime where absorption may be detectable at the highest energy bin observed with Fermi-LAT, while sources in the $\tau < 0.1$ regime are represented by grey markers. Different markers indicate the quality of the redshift determination: robust (spectroscopic) redshifts are shown as filled circles; lower-limit redshifts are represented by right arrows, with the reported redshift value corresponding to the center of the arrow; and uncertain redshifts (photometric or unknown quality) are represented by blue crosses. Approximately one-third of the 1CGH sources lack redshift assignments and are therefore not included in this plot. Note: the $\rm \tau_{(E,z)} $ values regarding the largest energy bin are listed in the 1CGH catalogue to help identify interesting targets}.
    \label{fig:Ebin}
\end{figure*}

To measure the EBL optical depth, one needs to build the $\gamma$-ray spectrum and compare the observed flux with the expected flux as a function of energy. For sources observed with Fermi-LAT, the $\gamma$-ray spectrum typically spans from tens of MeV to hundreds of GeV, and the flux can only be resolved for a few energy bins ($\rm E^{bin}$) (i.e. due to sensitivity limitations, Fermi-LAT usually cannot achieve high spectral resolution).

If we look for the most interesting sources for studying the cosmic gamma-ray horizon, we should focus on estimating the largest energy bin ($\rm E^{bin}_{max}$) where a source can still be detected. Any source with at least one energy bin that experiences a detectable degree of EBL attenuation is relevant to measure $\rm \tau_{(E,z)}$ values.

To emphasize sources that could be used to measure $\rm \tau_{(E,z)}$, we adopt a selection method based on the (z,$\rm E^{bin}_{max}$) position in the energy versus redshift plane. We calculate $\rm E^{bin}_{max}$ as the mean energy of the four highest-energy photons associated with a source, to create a robust estimate of the highest-energy bin. A similar requirement for the minimum photon counts was applied in the 3FHL catalogue \citep{2017ApJS..232...18A} to set a lower acceptance level for robust source detections. When calculating $\rm E^{bin}_{max}$, we use the clean photon sample (as described in Section \ref{sec:clean_photons}), considering source-type events and excluding PSF0 events (i.e., \texttt{evclass=128} and \texttt{evtype=56}).

In Figure \ref{fig:Ebin}, we illustrate the energy versus redshift relation and use the Saldana-Lopez EBL model \cite{2021MNRAS.507.5144S} as a reference framework to identify sources where gamma-ray absorption might be detectable. The Saldana-Lopez model offers an empirical determination of the evolving EBL spectrum and its uncertainties up to z$\sim$6; it is based on galaxy counts and multifrequency data (from UV to far-IR) covering the Hubble Space Telescope's `Cosmic Assembly Near-infrared Deep Extragalactic Legacy Survey' (CANDELS), and was designed to minimise uncertainties especially at higher redshift \citep{2011ApJS..197...35G, 2011ApJS..197...36K}.

However, our ultimate goal is to highlight a sample in which the $\tau_{(E,z)}$ values can be derived from the data, independently of any specific EBL model. The estimates of \cite{2021MNRAS.507.5144S} are similar to those by \cite{2010ApJ...712..238F} and both are representative of the most severe attenuation models among the ones considered. 

We used an EBL optical depth of $\rm \tau_{(E,z)} = 0.1$ as the lower limit to introduce an absorption flag (ABS-flag), indicating whether spectral absorption might be detectable at the highest energy bins. For cases where $\rm E^{bin}_{max}$ lies in the $\rm \tau_{(E,z)} > 0.1$ regime, the largest energy bin observed with Fermi-LAT likely experiences a detectable absorption greater than 10\% (i.e., >($\rm 1 - e^{-0.1}$)). These sources can be used to investigate the opacity along redshift, and have ABS-flag set to `1'.  

For sources with $\rm E^{bin}_{max}$ in the $\rm \tau_{(E,z)} < 0.1$ regime, the Fermi-LAT spectrum is likely unaffected by EBL attenuation, and these sources have an ABS-flag set to `0' in the 1CGH catalogue. These sources can be used to study the intrinsic $\gamma$-ray spectrum from blazars, helping to refine assumptions regarding their intrinsic flux. 

The $\rm \tau_{(E,z)}$ values --calculated using the Saldana-Lopez EBL model at $\rm E^{bin}_{max}$  \cite{2021MNRAS.507.5144S}-- are listed in the column `$\rm \tau_{Ebin}$'. Furthermore, the absorption fraction ($\rm 1 - e^{-\tau_{(E,z)}}$) is recorded in the `$\rm Abs_{Ebin}$' column, which can be used to create subsamples based on different absorption thresholds. In particular, for sources with multiple energy bins that experience detectable absorption, a single source can provide $\rm \tau_{(E,z)}$ estimates for multiple energy levels at a given redshift.

\subsection{Best Candidates for Optical Observation Campaigns}
\label{sec:opt_candidates} 

Following our extensive review of redshift information (sec. \ref{sec:redshift}), Appendix \ref{appendix:tableOptCand} highlights the high galactic latitude sources (|b| > 10$^{\circ}$) that are currently the best candidates for optical observations. We have identified two main types of sources to prioritize, which will further improve the catalogue and the science described above:

\begin{itemize}
    \item 1CGH sources that currently lack redshift information (z-flag = 0) but already have an optical or radio association are prioritised based on $\rm E^{bin}_{max}$, the highest energy bin detected with Fermi-LAT. These sources are particularly significant due to the presence of a clear counterpart for optical follow-up and their potential proximity to the $\rm \tau \sim 1$ horizon. Table \ref{tab:zflag_0}1 highlights 42 of these cases.   
    
    \item 1CGH sources with lower limit redshifts or photometric/uncertain redshifts (z-flag = 2 or 3) are prioritised based on their calculated opacity ($\rm \tau_{Ebin}$), using $\rm E^{bin}_{max}$ values. These sources could be experiencing significant $\gamma$-ray absorption, potentially detectable across multiple energy bins, and thus could provide valuable $\rm \tau{_{(E)}}$ measurements across different energies at a given redshift. Table \ref{tab:zflag_23}2 highlight 43 of these cases.
\end{itemize}

Any candidate for follow-up could also be considered in light of the redshift upper limits proposed by \cite{2024MNRAS.527.4763D}, which use EBL attenuation to constrain the distance of $\gamma$-ray blazars.

\section{Summary and Conclusions}

In this work, we presented the First Cosmic Gamma-ray Horizon (1CGH) catalogue, which lists all-sky Fermi-LAT detections above 10 GeV, compiled from 16 years of observations. The catalogue includes 2791 sources, with 62 representing new $\gamma$-ray detections, and to date, it represents a valuable sample to study the transparency of the Universe to VHE photons.

We significantly improved the redshift characterization of our catalogue by incorporating new spectroscopic redshift estimates whenever possible. By meticulously reviewing nearly 70 dedicated observational campaigns, we significantly reduced the redshift knowledge gap, introducing a z-flag system to categorise redshift reliability. 

The 1CGH catalogue highlights sources experiencing varying degrees of $\gamma$-ray attenuation and will contribute to a better determination of the EBL's density and its evolution over cosmic time. Using the Saldana-Lopez EBL model as a reference framework, we identified sources with $\gamma$-ray absorption above the $\rm \tau_{(E,z)} > 0.1$ limit, creating a targeted sample where absorption is likely measurable.

In addition, we introduced a complementary data exploration method by calculating the mean energy of the four highest-energy photons $\rm E^{bin}_{max}$, rather than relying solely on the highest-energy event. This provides a more stable and statistically reliable measure of the characteristic `largest energy level' observed for each source.

We identified sources that should be prioritised for follow-up optical observation campaigns. These include: (i) sources that lack redshift information but have a clear optical or radio association, prioritised based on the highest $\rm E^{bin}_{max}$ values; and (ii) sources with lower limit or uncertain redshift estimates that could benefit from further spectroscopic exploration, prioritised according to their $\tau$ values measured at $\rm E^{bin}_{max}$. These targets will not only help refine the redshift description of the catalogue but also improve the accuracy of EBL measurements at higher redshift, thus reducing uncertainties in SFR density determination during the EoR, at z$\sim$6-7.

Following our redshift review (Section \ref{sec:redshift}), approximately 72\% of the 1CGH sources still lack robust redshift determination. This significant gap poses challenges for the very high-energy community as a whole, and calls attention to the need for extensive follow-up optical observation campaigns targeting $\gamma$-ray blazars.  

\section*{Acknowledgements}

The authors thank the anonymous Referee for the careful review and helpful comments that improved the quality -and impact- of our work. BA thanks the Institute of Astrophysics and Space Sciences (IA) at the University of Lisbon for their ongoing support. BA is currently a Marie Sk\l{}odowska-Curie Postdoctoral Fellow at IA, funded by the European Union's Horizon 2020 research and innovation program under the MSCA agreement No. 101066981. BA also acknowledges the support from `Fundação para a Ciência e a Tecnologia' (FCT) through the research grant UIDP/04434/2020 (DOI: 10.54499/UIDP/04434/2020) in support of general IA activities. This work was produced with the support of the "National Distributed Computing Infrastructure" (INCD), Portugal (\url{https://www.incd.pt/}), funded by FCT and FEDER under project 01/SAICT/2016 No. 022153. We acknowledge FCT's support through the A1 Computation Call (project's doi: 10.54499/2023.09549.CPCA.A1) for the allocation of HPC resources at INCD. We are grateful to the entire {\it Fermi}-LAT collaboration for maintaining a publicly accessible mission database, enabling discoveries across the entire $\gamma$-ray community. We also used archival data and bibliographic information from the NASA-IPAC Extragalactic Database (NED), and data/software facilities maintained by the Space Science Data Center (SSDC) from the Italian Space Agency. We extend our thanks to the TeVCat team, especially Deirdre Horan and Scott Wakely, for their continuous work on VHE detections, as well as NASA and NSF for their support of these follow-up efforts.


\section*{Data Availability}

All catalogues used to build the sample of $\gamma$-ray candidates are publicly available and cited within this work. The 1CGH catalogue will be made publicly available through VizieR at \url{https://vizier.cds.unistra.fr/}, and GitHub. The Fermi-LAT database and the Science Tools used to build the 1CGH catalogue are also publicly accessible at \url{https://fermi.gsfc.nasa.gov/ssc/data/access/}.



\bibliographystyle{mnras}
\bibliography{bibliograph} 



\appendix

\section{Column Descriptions of the 1CGH Catalogue}
\label{appendix:1CGHcolumns} 

The 1CGH catalogue contains detailed metadata for 2791 sources detected above 10 GeV based on 16 years of Fermi-LAT observations. Table \ref{tab:1CGHcolumns} provides a description of each column and details its content.


\begin{table*}
\centering
\caption{Column descriptions of the 1CGH catalogue. The table lists the column names, units where applicable, and detailed explanations of their contents.}
\label{tab:1CGHcolumns}
\begin{tabular}{|l|p{14cm}|}
\hline
\textbf{Column Name}       & \textbf{Description} \\
\hline
1CGH\_name                 & Source identifier, formatted as \texttt{1CGHJ0123+0123}, based on J2000 coordinates. Includes four digits each for R.A. and Dec. For associated sources, the coordinates are from the counterpart; for unassociated sources, we follow 4FGL-DR4 astrometry. \\
RA\_1CGH                   & Right Ascension (J2000) in degrees. \\
DEC\_1CGH                  & Declination (J2000) in degrees. \\
Counterpart\_name          & Name of the source from catalogues such as 5BZcat, 3HSP, 4FGL-DR4, or others. \\
N0[1E-16]                  & Normalization constant of the power-law fit in units of photons/cm$^2$/s/MeV at the pivot energy. \\
N0\_err[1E-16]             & Uncertainty in the normalization constant. \\
Index                      & Photon spectral index of the power-law model. \\
Index\_err                 & Uncertainty in the photon spectral index. \\
Flux[1E-12]                & Integrated photon flux (10–800 GeV) in units of photons/cm$^2$/s, estimated with a power-law fit.  \\
Flux\_err[1E-12]           & Uncertainty in the integrated photon flux. \\
TS                         & Test Statistic value for the source detection. \\
sigma                      & The significance for the source detection. \\ 
z                          & Redshift of the source. \\
zflag                      & Redshift reliability flag: (1) spectroscopic, (2) photometric/uncertain, (3) lower limit. \\
z\_origin                  & Reference for the redshift, as described in the text. For cases with no corresponding redshift value, it tracks sources where spectroscopic observation has been attempted. \\
Nph\_128                   & Number of photons associated (Source\_type events, \texttt{evclass=128}). \\
Emax\_128[GeV]             & Maximum photon energy detected (\texttt{evclass=128}). \\
Nph\_512                   & Number of photons associated (UltraClean\_type events, \texttt{evclass=512}). \\
Emax\_512[GeV]             & Maximum photon energy detected (\texttt{evclass=512}). \\
Ebin\_max[GeV]             & Mean energy of the four highest-energy photons (\texttt{evclass=128}). \\
TransmissionEbin           & Flux transmission factor at $\rm E^{bin}_{max}$. \\
AbsEbin                    & Flux absorption factor at $\rm E^{bin}_{max}$, calculated as ($1 -$ Transmission). \\
tauEbin                    & EBL optical depth at $\rm E^{bin}_{max}$. \\
Abs\_flag                  & Absorption flag: (1) $\tau > 0.1$ (moderate to severe absorption), (0) $\tau \leq 0.1$. \\
GAL\_LAT                   & Galactic latitude in degrees. \\
BZcat                      & Identifier in the 5BZcat catalogue, if available. \\
z\_3HSP                    & Redshift from the 3HSP catalogue, if available. \\
zflag\_3HSP                & Redshift flag from the 3HSP catalogue: (1) spectroscopic, (2) uncertain, (3) lower limit redshift, (4) photometric -with featureless optical spectrum, (5) photometric -without optical spectrum. \\
4FGL\_counter\_name        & Counterpart name in 4FGL-DR4. \\
TEVCAT\_FLAG\_4FGLdr4      & Flag for TeVCat association, as per 4FGL-DR4. \\
ASSOC\_TEV\_4FGLdr4        & TeV association in 4FGL-DR4. \\
CLASS1\_4FGLdr4            & Classification in 4FGL-DR4. \\
ASSOC1\_4FGLdr4            & Counterpart name in 4FGL-DR4. \\
Name\_4LACdr3              & Counterpart name in 4LAC-DR3. \\
z\_4LACdr3\_1              & Redshift as reported in 4LAC-DR3. \\
Name\_3FHL                 & Counterpart name in 3FHL. \\
z\_3FHL                    & Redshift as reported in 3FHL. \\
Curvature\_3FHL   & Detectable spectral curvature in 3FHL, (1)yes, (2)no, (-) unknown. Sources marked as "unknown" are out of the 3FHL catalogue. \\
\hline
\end{tabular}
\end{table*}

\section{Long Tables: Best Candidates for optical Observations}
\label{appendix:tableOptCand}

In this section, we present tables highlighting the best candidates for optical follow-up observations from the 1CGH catalogue. These sources have been selected based on their potential to provide insights about the CGH. Two categories of sources are prioritized: Table \ref{tab:zflag_0}1 those without redshift information but with clear optical, infrared, or radio associations, and Table \ref{tab:zflag_23}2 those with uncertain or lower-limit redshifts where significant $\gamma$-ray absorption is predicted. These candidates will refine the redshift completeness of the 1CGH catalogue. Each source selection is based -respectively- on its highest-energy bin ($\rm E^{bin}_{max}$), and on its calculated optical depth $\rm \tau_{(Ebin)}$, as described in sec. \ref{sec:opt_candidates}. Table \ref{tab:zflag_0} provides the IR counterparts from the `Wide-field Infrared Survey Explorer' AllWISE catalogue \citep{Cutri2013,2014yCat.2328....0C} which are meant to help guide optical observations\footnote{For 4FGLJ0351.2-6103, the AllWISE IR counterpart is tentative; and for 4FGLJ2127.6-5959, a tentative IR counterpart is close by NGC7059.}.

\begin{table*}
\label{tab:zflag_0}
\caption{This table lists the sources in the catalogue without redshift information, ordered by their $\rm E^{bin}_{max}$, and where we could retrieve clear IR/Optical counterparts for observation. The first three columns show the source names, right ascension (R.A.), and declination (Dec.) in degrees (J2000), respectively. The column `4FGL-DR4' and `ASSOC1-4FGL' shows the source and counterpart names reported in 4FGL-DR4 catalogue. Next, the column `TS' provides the Test Statistic value derived for 1CGH sources detected above 10 GeV, integrating over 16 years of Fermi-LAT observations. The column `$\rm E_{max}^{bin}$' list the highest energy bin detected. The column `Rmag' provides the R band magnitude retrieved using SSDC Sky Explorer.(\url{https://tools.ssdc.asi.it/}). The last column, `AllWISE' provides the IR counterpart from the AllWISE catalogue (cases marked with a `*' sign represent a tentative association and should be considered with care). For sources whose 'Name' begins with 4FGLJ, the R.A. and Dec. correspond to the counterpart position \citep[`ASSOC1',][]{2023arXiv230712546B}.}  
{\def\arraystretch{1.3}
 \begin{tabular}{llrcccrrl}
\hline
Name  & R.A.(deg)  &  Dec.(deg)  &  4FGL-DR4 & ASSOC1-4FGL  &  TS  & $\rm E^{bin}_{max}$ & Rmag  &  AllWISE \\
 \hline
  4FGLJ1634.9+1222        &   248.76049  &     12.3619         & J1634.9+1222    &  NVSSJ163502+122142         &      15 &   233  & 20.04 &  J163502.41+122142.1     \\     
  3HSPJ013632.6+390559    &    24.13579  &     39.0997         & J0136.5+3906    &  B30133+388                 &    4708 &   183  & 15.86 &  J013632.59+390559.1     \\     
  3HSPJ130738.0$-$425938  &   196.90825  &    $-$42.9941       & J1307.6$-$4259  &  1RXSJ130737.8$-$425940     &     732 &   150  & 16.26 &  J130737.98$-$425938.9   \\     
  4FGLJ0840.1$-$0225      &   130.06121  &     $-$2.4531       & J0840.1$-$0225  &  PMNJ0840$-$0227            &      37 &   149  & 14.57 &  J084014.69$-$022711.3   \\     
  4FGLJ1334.1$-$3521      &   203.55010  &    $-$35.3372       & J1334.1$-$3521  &  PKS1331$-$350              &      21 &   138  & 12.25 &  J133412.03$-$352014.2   \\     
  3HSPJ074642.0$-$475455  &   116.67629  &    $-$47.9152       & J0746.6$-$4754  &  PMNJ0746$-$4755            &    1223 &   134  & 18.18 &  J074642.31$-$475455.0   \\     
  3HSPJ230436.7+370507    &   346.15295  &     37.0854         & J2304.6+3704    &  1RXSJ230437.1+370506       &     297 &   98   & 20.2  &  J230436.71+370507.4     \\     
  3HSPJ102634.4$-$854314  &   156.64316  &    $-$85.7206       & J1027.0$-$8542  &  PKS1029$-$85               &     528 &   95   & 17.27 &  J102634.36$-$854314.2   \\     
  5BZBJ0700$-$6610        &   105.13017  &    $-$66.1792       & J0700.5$-$6610  &  PKS0700$-$661              &     986 &   94   & 15.29 &  J070031.25$-$661045.2   \\     
  3HSPJ230940.8$-$363248  &   347.42012  &    $-$36.5468       & J2309.7$-$3632  &  WISEJ230940.84$-$363248    &     122 &   87   & 18.62 &  J230940.84$-$363248.7   \\     
  3HSPJ033349.0+291631    &    53.45416  &     29.2754         & J0333.7+2916    &  TXS0330+291                &     591 &   85   & 19.1  &  J033349.00+291631.5     \\     
  5BZBJ1849$-$4314        &   282.358    &    $-$43.2369       & J1849.4$-$4313  &  PMNJ1849$-$4314            &     121 &   82   & 17.8  &  J184925.92$-$431413.3   \\     
  4FGLJ0703.2$-$3914      &   105.80268  &    $-$39.2385       & J0703.2$-$3914  &  1RXSJ070312.7$-$391417     &      52 &   81   & 17.6  &  J070312.65$-$391418.8   \\     
  4FGLJ1608.0$-$2038      &   241.98719  &    $-$20.6617       & J1608.0$-$2038  &  NVSSJ160756$-$203942       &      50 &   80   & 18.22 &  J160756.90$-$203943.5   \\     
  3HSPJ130421.0$-$435310  &   196.0875   &    $-$43.8861       & J1304.3$-$4353  &  1RXSJ130421.2$-$435308     &    1049 &   77   & 17.48 &  J130421.00$-$435310.2   \\     
  4FGLJ2345.9+3413        &   356.47985  &     34.2342         & J2345.9+3413    &  1RXSJ234554.2+341419       &      21 &   74   & 17.8  &  J234555.21+341404.5     \\     
  4FGLJ2143.9+3337        &   325.95889  &     33.6196         & J2143.9+3337    &  MG3J214351+3337            &      74 &   73   & 20.6  &  *J214350.20+333711.7  \\     
  4FGLJ1345.6$-$3356      &   206.42935  &    $-$33.9453       & J1345.6$-$3356  &  NVSSJ134543$-$335643       &      76 &   73   & 19.58 &  J134543.04$-$335643.3   \\     
  5BZBJ2108$-$6637        &   317.21592  &    $-$66.6229       & J2108.9$-$6638  &  PKS2104$-$668              &     198 &   73   & 18.22 &  J210851.80$-$663722.7   \\     
  4FGLJ0836.0$-$8015      &   128.95466  &    $-$80.2696       & J0836.0$-$8015  &  2MASSJ08354940$-$8016114   &      29 &   69   & 20.74 &  J083549.36$-$801610.8   \\     
  4FGLJ2300.8$-$0736      &   345.22798  &     $-$7.5954       & J2300.8$-$0736  &  2MASSJ23005469$-$0735438   &      34 &   69   & 19.51 &  J230054.70$-$073543.5   \\     
  3HSPJ081012.0$-$703047  &   122.55016  &    $-$70.5130       & J0809.9$-$7028  &  SUMSSJ081011$-$703048      &      12 &   68   & 19.44 &  J081012.04$-$703047.1   \\     
  3HSPJ104756.9$-$373730  &   161.98725  &    $-$37.6252       & J1047.9$-$3738  &  GALEXASCJ104756$-$373730   &      45 &   65   & 19.7  &  J104756.94$-$373730.8   \\     
  4FGLJ1804.4+5249        &   271.09484  &     52.8205         & J1804.4+5249    &  6CB180317.2+524912         &      14 &   65   & 15.87 &  J180422.64+524915.2     \\     
  3HSPJ195547.9+021512    &   298.94941  &      2.2535         & J1955.7+0214    &  NVSSJ195547+021514         &      92 &   63   & 20.95 &  J195547.95+021517.9     \\     
  4FGLJ1836.5+1948        &   279.13380  &     19.8461         & J1836.5+1948    &  NVSSJ183632+195047         &      54 &   62   & 18.83 &  J183632.11+195046.3     \\     
  5BZBJ0718$-$4319        &   109.68183  &    $-$43.3304       & J0718.6$-$4319  &  PMNJ0718$-$4319            &     760 &   62   & 17.73 &  J071843.63$-$431949.7   \\     
  4FGLJ2025.3$-$2231      &   306.31322  &    $-$22.5051       & J2025.3$-$2231  &  NVSSJ202515$-$223016       &      55 &   62   & 18.15 &  J202515.17$-$223018.4   \\     
  4FGLJ2127.6$-$5959      &   321.87039  &    $-$60.0137       & J2127.6$-$5959  &  NGC7059(?)                 &      16 &   60   & 17.79 & *J212729.34$-$600055.8   \\    
  4FGLJ0624.2$-$2943      &    96.09291  &    $-$29.7469       & J0624.2$-$2943  &  1RXSJ062422.3$-$294449     &      19 &   59   & 21.46 &  J062422.49$-$294446.7   \\     
  4FGLJ2049.0$-$4020      &   312.29171  &    $-$40.3419       & J2049.0$-$4020  &  PKS2045$-$405              &      19 &   59   & 19.53 &  J204910.02$-$402030.9   \\     
  3HSPJ151444.0$-$772254  &   228.68341  &    $-$77.3817       & J1514.4$-$7719  &  1RXSJ151448.8$-$772249     &      30 &   58   & 19.74 &  J151444.02$-$772254.2   \\     
  4FGLJ0351.2$-$6103      &    57.75249  &    $-$61.0467       & J0351.2$-$6103  &  SUMSSJ035100$-$610248      &      15 &   57   & 16.61 & *J035059.83$-$610241     \\         
  4FGLJ0928.7$-$3529      &   142.20762  &    $-$35.4969       & J0928.7$-$3529  &  NVSSJ092849$-$352947       &      15 &   56   & 19.64 &  J092849.82$-$352948.8   \\     
  3HSPJ145543.7$-$760052  &   223.93204  &    $-$76.0145       & J1455.8$-$7601  &  SUMSSJ145543$-$760054      &      45 &   55   & 18.41 &  J145543.69$-$760052.2   \\     
  4FGLJ0942.7$-$2028      &   145.65761  &    $-$20.4536       & J0942.7$-$2028  &  1RXSJ094237.9$-$202720     &      75 &   55   & 19.96 &  J094237.81$-$202713.2   \\     
  3HSPJ094709.5$-$254100  &   146.78966  &    $-$25.6833       & J0947.1$-$2541  &  1RXSJ094709.2$-$254056     &     397 &   55   & 18.16 &  J094709.52$-$254059.9   \\     
  4FGLJ1513.4$-$3721      &   228.32775  &    $-$37.3365       & J1513.4$-$3721  &  2MASSJ15131867$-$3720114   &      56 &   54   & 17.83 &  J151318.66$-$372011.5   \\     
  3HSPJ080215.9$-$094210  &   120.56625  &     $-$9.7030       & J0802.3$-$0942  &  WISEJ080215.63$-$0942(?)   &     235 &   54   & 19.04 &  J080215.90$-$094210.9   \\     
  4FGLJ1828.0+2634        &   276.98555  &     26.5535         & J1828.0+2634    &  NVSSJ182756+263313         &      40 &   50   & 19.4  &  J182756.54+263313.2     \\     
  3HSPJ213004.8$-$563222  &   322.51987  &    $-$56.5394       & $-$             &   $-$                       &      18 &   50   & 20.48 &  J182756.54+263313.2     \\     
  3HSPJ134706.9$-$295842  &   206.77866  &    $-$29.9784       & J1347.1$-$2959  &  NVSSJ134706$-$295840       &      62 &   50   & 18.8  &  J134706.88$-$295842.4   \\             
 \end{tabular}
}
\end{table*}

\begin{table*}
\label{tab:zflag_23}
\caption{This table lists the sources in the catalogue with uncertain redshift information, ordered by their $\tau_{(E,z)}$ values calculated at $\rm E^{bin}_{max}$. The first columns show the source names, R.A., and Dec. in degrees (J2000), respectively. The fourth column presents redshift information from the literature, with a `(?)' flag indicating uncertain or photometric values and a `>' symbol denoting lower limits. The `z-origin' column specifies the literature reference for the redshift, using the short names described in Section \ref{sec:redshift}. The `b(deg)' column shows the galactic latitude in degrees, followed by the associated `4FGL-DR4' name. The `TS' column provides the Test Statistic value derived for 1CGH sources detected above 10 GeV, integrating over 16 years of Fermi-LAT observations. The columns `$\rm E_{max}^{bin}$' and `$\rm \tau_{Ebin}$' list the largest energy bin detected and the corresponding EBL optical depth $\tau_{(E,z)}$, respectively. For sources whose Name' begins with 4FGLJ, the R.A. and Dec. correspond to the associated counterpart position\citep[see the `ASSOC1-4FGL' column in,][]{2023arXiv230712546B}.} 
{\def\arraystretch{1.3}
 \begin{tabular}{llrccccrrrc}
\hline
Name  & R.A. (deg)  &  Dec. (deg)  &  z  &  z-origin & b(deg)  &  4FGL-DR4 & TS &  $\rm E^{bin}_{max} $ &  $\rm \tau_{Ebin}$ \\  
 \hline
  3HSPJ224753.2+441315   &   341.97175  &    44.22097   &   1.9(?)     &  Foschini22   &   $-$13.2   &   J2247.8+4413   &     375 &   179  &  5.08  \\    
  3HSPJ015307.4+751742   &    28.28070  &    75.29522   &   2.35(?)    &  Foschini22   &    12.9     &   J0153.0+7517   &     168 &   149  &  4.94  \\  
  4FGLJ0400.7+3920       &    60.18911  &    39.35271   &   1.1(?)     &  Foschini22   &   $-$10.2   &   J0400.7+3920   &      18 &   209  &  3.34  \\  
  3HSPJ200506.0+700439   &   301.27487  &    70.07763   &   2.32(?)    &  Foschini22   &    19.4     &   J2005.1+7003   &     311 &    66  &  1.60  \\  
  3HSPJ052542.4$-$601340 &    81.42675  &   $-$60.22783 &   1.78(?)    &  Kaur16       &   $-$33.8   &   J0525.6$-$6013 &     249 &    70  &  1.21  \\  
  4FGLJ1937.0+8354       &   294.41522  &    83.94139   &   1.94(?)    &  Foschini22   &    25.7     &   J1937.0+8354   &      39 &    61  &  1.12  \\  
  4FGLJ1858.1+7318       &   284.58486  &    73.28700   &   0.471(?)   &  Foschini22   &    25.4     &   J1858.1+7318   &      43 &   221  &  1.10  \\  
  3HSPJ031612.7+090443   &    49.05304  &     9.07866   &   0.372(?)   &  3HSPcat      &   $-$39.5   &   J0316.2+0905   &     621 &   255  &  0.99  \\  
  3HSPJ142829.9+743002   &   217.12454  &    74.50061   &   0.31(?)    &  3HSPcat      &    41.0     &   J1428.8+7429   &      18 &   294  &  0.95  \\  
  3HSPJ144800.6+360831   &   222.00245  &    36.142     &   >0.738     &  Paiano20     &    63.7     &   J1448.0+3608   &    1085 &   137  &  0.94  \\  
  5BZBJ1248+5820         &   192.07825  &    58.34131   &   0.508(?)   &  Foschini22   &    58.7     &   J1248.3+5820   &    4149 &   189  &  0.94  \\  
  5BZBJ0007+4712         &     1.99988  &    47.20214   &   >1.659     &  Shaw13       &   $-$15.0   &   J0008.0+4711   &     709 &    61  &  0.87  \\  
  3HSPJ122337.0$-$303250 &   185.90420  &   $-$30.54725 &   >0.875     &  Desai19      &    31.9     &   J1223.6$-$3032 &      76 &   113  &  0.87  \\  
  4FGLJ0318.7+2135       &    49.69028  &    21.57685   &   1.83(?)    &  Foschini22   &   $-$29.6   &   J0318.7+2135   &     171 &    55  &  0.85  \\  
  3HSPJ063059.5$-$240646 &    97.748    &   $-$24.11280 &   >1.239     &  3HSPcat      &   $-$14.9   &   J0630.9$-$2406 &    3589 &    78  &  0.80  \\  
  5BZBJ1918$-$4111       &   289.56687  &   $-$41.19189 &   >1.591     &  Shaw13       &   $-$22.2   &   J1918.2$-$4111 &     555 &    60  &  0.79  \\  
  5BZBJ1925$-$2219       &   291.41579  &   $-$22.32642 &   1.35(?)    &  Foschini22   &   $-$17.1   &   J1925.8$-$2220 &      37 &    69  &  0.76  \\  
  5BZBJ2300+3137         &   345.09512  &    31.61789   &   >1.498     &  Shaw13       &   $-$25.5   &   J2300.3+3136   &     283 &    62  &  0.76  \\  
  5BZBJ1314+2348         &   198.68254  &    23.80744   &   2.053(?)   &  SDSSdr18     &    83.7     &   J1314.7+2348   &     387 &    46  &  0.76  \\  
  4FGLJ0453.1+6322       &    73.30189  &    63.35496   &   2.1(?)     &  Foschini22   &    12.1     &   J0453.1+6322   &      39 &    45  &  0.75  \\  
  3HSPJ154015.1+815505   &   235.06625  &    81.91822   &   >0.67      &  Paiano17     &    32.9     &   J1540.1+8155   &     938 &   129  &  0.72  \\  
  3C 66A                 &    35.67334  &    43.04319   &   >0.3347    &  Furniss13    &   $-$16.7   &   J0222.6+4302   &    7060 &   230  &  0.70  \\   
  5BZBJ0856$-$1105       &   134.17421  &   $-$11.08733 &   >1.397     &  Shaw13       &    21.4     &   J0856.6$-$1105 &     517 &    64  &  0.70  \\  
  5BZBJ0612+4122         &    93.21325  &    41.37706   &   >1.107     &  Paiano20     &    10.9     &   J0612.8+4122   &    1383 &    80  &  0.68  \\  
  3HSPJ181118.0+034113   &   272.82508  &     3.68711   &   0.717(?)   &  Foschini22   &    10.6     &   J1811.3+0340   &     473 &   117  &  0.67  \\  
  3HSPJ043145.1+740326   &    67.93775  &    74.05738   &   1.35(?)    &  Foschini22   &    17.3     &   J0431.8+7403   &     226 &    65  &  0.67  \\  
  3HSPJ103744.3+571155   &   159.43458  &    57.19880   &   >0.62      &  Meisner10    &    51.7     &   J1037.7+5711   &    3071 &   131  &  0.65  \\  
  3HSPJ003514.1+151504   &     8.81129  &    15.25116   &   >0.64      &  3HSPcat      &   $-$47.4   &   J0035.2+1514   &     485 &   122  &  0.60  \\  
  3HSPJ112048.1+421212   &   170.20025  &    42.2035    &   >0.35      &  3HSPcat      &    66.1     &   J1120.8+4212   &    1811 &   199  &  0.58  \\  
  S5 0716+714            &   110.47250  &    71.34332   &   0.31(?)    &  Foschini22   &    28.0     &   J0721.9+7120   &   11282 &   218  &  0.57  \\  
  3HSPJ170433.8$-$052840 &   256.14095  &    $-$5.47797 &   >0.7       &  Paiano17     &    20.7     &   J1704.5$-$0527 &     244 &   109  &  0.55  \\  
  3HSPJ152048.9$-$034851 &   230.20370  &    $-$3.81433 &   >0.868     &  Goldini20    &    42.5     &   J1520.8$-$0348 &     421 &    88  &  0.53  \\  
  3HSPJ193419.6+600139   &   293.58175  &    60.02763   &   1.38(?)    &  Foschini22   &    18.1     &   J1934.2+6002   &      44 &    56  &  0.52  \\  
  4FGLJ1304.2$-$2412     &   196.06955  &   $-$24.20465 &   1.26(?)    &  Foschini22   &    38.5     &   J1304.2$-$2412 &     155 &    60  &  0.51  \\  
  5BZBJ1312$-$2156       &   198.1315   &   $-$21.93983 &   >1.485     &  Shaw13       &    40.6     &   J1312.4$-$2156 &     358 &    50  &  0.50  \\  
  4FGLJ0028.5+2001       &     7.12424  &    20.00742   &   1.552(?)   &  4LACdr3      &   $-$42.5   &   J0028.5+2001   &     489 &    48  &  0.50  \\  
  3HSPJ175615.1+552218   &   269.06625  &    55.37166   &   >0.657     &  3HSPcat      &    29.7     &   J1756.3+5522   &     264 &   106  &  0.47  \\  
  4FGLJ0355.3+3909       &    58.81912  &    39.15272   &   0.846(?)   &  Foschini22   &   $-$11.0   &   J0355.3+3909   &      18 &    85  &  0.47  \\  
  3HSPJ210421.9$-$021238 &   316.09137  &    $-$2.21080 &   >0.45      &  3HSPcat      &   $-$30.3   &   J2104.3$-$0212 &      88 &   140  &  0.43  \\  
  3HSPJ222129.3$-$522527 &   335.37208  &    $-$52.4243 &   >0.34      &  3HSPcat      &   $-$52.3   &   J2221.5$-$5225 &     426 &   174  &  0.42  \\  
  3HSPJ125359.3+624257   &   193.49725  &     62.716    &   >0.867     &  Shaw13       &    54.4     &   J1253.8+6242   &     179 &    79  &  0.42  \\  
  4FGLJ1537.9$-$1344     &   234.48793  &    $-$13.7262 &   0.984(?)   &  Foschini22   &    32.5     &   J1537.9$-$1344 &      31 &    70  &  0.42  \\  
  5BZBJ2016$-$0903       &   304.10013  &     $-$9.0592 &   >0.605     &  Shaw13       &   $-$22.9   &   J2016.3$-$0903 &     286 &   108  &  0.42  \\  
 \end{tabular}
}
\end{table*}


\bsp	
\label{lastpage}
\end{document}